\documentclass[12pt,preprint]{aastex}

\usepackage{natbib}
\usepackage{multirow}
\usepackage{pdfpages}
\usepackage{graphicx}
\usepackage{fancyvrb}
\usepackage{epstopdf}
\usepackage[mathscr]{euscript}

\slugcomment{}

\shorttitle{}
\shortauthors{Osten et al.}

\newcommand{\Ufilt}{$U$}
\newcommand{\ufilt}{$u$}
\newcommand{\uvwfilt}{$uvw2$}
\newcommand{\Bfilt}{$B$}
\newcommand{\bfilt}{$b$}
\newcommand{\Vfilt}{$V$}
\newcommand{\vfilt}{$v$}
\newcommand{\Rfilt}{$R$}
\newcommand{\Ifilt}{$I$}
\VerbatimFootnotes

\begin{document}
\title{A Very Bright, Very Hot, and Very Long Flaring Event from the M Dwarf Binary System DG CVn}
\shorttitle{The Flare That Kept on Going}
 
\author{Rachel A. Osten\altaffilmark{1}}
\affil{Space Telescope Science Institute}
\email{osten@stsci.edu}
\author{Adam Kowalski}
\affil{ U. Md/GSFC}
\author{Stephen A. Drake}
\affil{ USRA/CRESST and NASA/GSFC}
\author{Hans Krimm}
\affil{ USRA/CRESST}
\author{Kim Page}
\affil{ X-ray and Observational Astronomy Group, Department of Physics \&
Astronomy, University of Leicester, Leicester, LE1 7RH, UK}
\author{Kosmas Gazeas}
\affil{ Department of Astrophysics, Astronomy and Mechanics, University of Athens, GR-15784 Zografos, Athens, Greece}
\author{Jamie Kennea}
\affil{ Penn State}
\author{Samantha Oates}
\affil{ Instituto de Astrofísica de Andaluc\'{i}a (IAA-CSIC), Glorieta de la Astronom\'{i}a s/n, E-18008, Granada, Spain }
\author{Mathew Page}
\affil{ UCL}
\author{Enrique de Miguel\altaffilmark{2}}
\affil{Departamento de Fisica Aplicada, Facultad de Ciencias Experimentales, Universidad de Huelva, 21071 Huelva, Spain}
\author{Rudolf Nov\'{a}k}
\affil{Research Centre for Toxic Compounds in the Environment, Faculty of Science, Masaryk University, 
Kamenice 3, 625 00 Brno, Czech Republic}
\author{Tomas Apeltauer}
\affil{Brno University of Technology, Faculty of Civil Engineering, Veveri 331/95, 602 00 Brno, Czech Republic}
\author{Neil Gehrels}
\affil{NASA/GSFC}

\altaffiltext{1}{Also at Center for Astrophysical Sciences, Johns Hopkins University, Baltimore, MD 21218}
\altaffiltext{2}{Also CBA-Huelva, Observatorio del CIECEM, Parque Dunar Matalasca\~{n}as, 21760 Almonte, Huelva, Spain}

\begin{abstract}
On April 23, 2014, the Swift satellite responded to a hard X-ray transient detected by its
Burst Alert Telescope, which turned out to be a stellar flare from a nearby, young 
M dwarf binary DG~CVn. We utilize observations at X-ray, UV, optical, and radio wavelengths
to infer the properties of two large flares.
The X-ray spectrum of the primary outburst can be described over the
0.3-100 keV bandpass by either
a single very high temperature plasma or a nonthermal thick-target bremsstrahlung model, 
and we rule out the nonthermal model based on energetic grounds.
The temperatures were the highest seen spectroscopically 
in a stellar flare, at T$_{X}$ of 290 MK. 
The first event was followed by a comparably energetic event almost a day later.
We constrain the photospheric area involved in each of the two flares
to be $>$10$^{20}$ cm$^{2}$, and find evidence from flux ratios in the second
event of contributions to the white light
flare emission in addition to the usual hot, T$\sim$10$^{4}$K blackbody emission seen in the impulsive
phase of flares.
The radiated energy in X-rays and white light reveal these events to be the 
two most energetic X-ray flares observed from an M dwarf, with 
X-ray radiated energies in the 0.3-10 keV
bandpass of 4$\times$10$^{35}$ and 9$\times$10$^{35}$ erg, and 
optical flare energies at 
E$_{V}$ of 2.8$\times$10$^{34}$ and
5.2$\times$10$^{34}$ erg, respectively. 
The results presented here should be integrated into updated modelling of the astrophysical
impact of large stellar flares on close-in exoplanetary atmospheres.
\end{abstract}

\keywords{stars: flare, stars: coronae, stars: individual (DG CVn)}

\section{Introduction }

Most of what is known about the mechanisms producing stellar flares is
informed by the detailed observations of flares on the Sun. Solar flares occur in close proximity to active regions 
(ARs), which are effectively localized magnetic field regions of 1-2 kG strength. 
Loops from these ARs extend into the solar corona;
as the footpoints of these loops are jostled by solar convective motions, they are twisted and distorted until 
magnetic reconnection occurs near the loop tops \citep{parker1988,benzgudel2010}.
The reconnection event is accompanied by a sudden release of energy, resulting in 
the acceleration of electrons and ions in these loops up to MeV energies, which stream both towards and 
away from the Sun, emitting nonthermal radio (gyrosynchrotron) and X-ray emission (particularly at the loop footpoints) 
as they move \citep{dennis1989}. 
These energetic particles stream down to the loop footpoints and 
deposit substantial energy to the lower solar atmosphere (the chromosphere), “evaporating” and heating plasma from 
this region to fill the flaring loop(s) with plasma \citep{lin2011}.
In the “decay” phase of the flare, the thermal emission dominates the X-ray emission, 
although in some large solar flares a nonthermal X-ray component may persist as a continuous source of energy
\citep{kontar2008}.

Young stars and stars in close binary systems rotate much more rapidly than the Sun, and, in consequence, 
have much stronger levels of magnetic activity, i.e., greater coverage by starspots 
and ARs, stronger chromospheric and coronal emission, and more frequent and powerful flares \citep{meibom2007,morgan2016}.
There is a large disparity between the extremes of solar and stellar flares: while the largest solar
flares have radiated energies exceeding 10$^{32}$ erg, and maximum coronal temperatures
of a few tens of MK \citep{hotsolarflares}, large stellar flares can be 10$^{6}$ times more energetic, with
coronal temperatures around 100 MK \citep{osten2007} and large energetic releases up to 10$^{38}$ erg
\citep{cftucflare,osten2007}.
A 2008 flare of the nearby 30-300 Myr old M dwarf flare star EV Lac \citep{osten2010} had a lower limit on energetic release of 
6$\times$ 10$^{34}$ erg.
\citet{caramazza2007} found X-ray flares on very young low mass stars to range up to
2$\times$10$^{35}$ erg, and \citet{tsuboi2014} found flares from active binary
systems to range up to 10$^{38}$ erg.
The interpretation of these stellar flaring events assumes that the same physical processes are at work as
in the solar case, as confirmed by multiwavelength observations of plasma heating and
particle acceleration in stellar flares \citep{benzgudel2010}. The largest stellar flares,
with their extreme parameters of temperature and energy release, clearly test this correspondence.
Initial suppositions of a transition from solar-stellar flare scaling laws has come from the
work of \citet{getman2008}, but those data could not determine flare temperatures accurately.

DG CVn (GJ 3789) is an interesting, albeit poorly studied member of this class of nearby, very young low-mass stars. It 
is noted as having an unusually active chromosphere \citep{beers1994} and corona \citep{hunsch1999}, as 
well as being one of the brightest nearby stellar radio emitters \citep{helfand1999}.  Subsequent studies 
confirm that it exhibits optical flares and sub-day rotational modulation \citep{robb1994}, 
with a measured photospheric line broadening of 51 km/s \citep{mohantybasri2003}
indicative of a very short rotational 
period of $<$ 8 hours. DG~CVn is a binary, as revealed by the double-lined spectrum noted in \citet{gizis2002}. 
Adaptive optics imaging of DG CVn \citep{beuzit2004} reveal it to be a close (0.2$^{\prime\prime}$ separation) 
visual binary system, with two components of near-equal optical brightness ($\Delta$V $\sim$ 0.3) and spectral types of M4Ve. 
The distance to DG~CVn, from a large study of the trigonometric parallaxes and kinematics of nearby active stars,
is 18 pc, with a space motion consistent with the system being a member of the population of 30-Myr old stars in the
solar neighborhood \citep{riedel2014}.
They quote a combined systemic $\log L_{X}$ and $\log L_{X}/L_{\rm bol}$, from which (by dividing $L_{\rm bol}$ equally between the two
components) a luminosity $\log (L_{\rm bol}/L_{\odot})$=$-1.72$ is obtained.  \citet{mohantybasri2003} determine a system T$_{\rm eff}$ of
3175 K, which combining with $L_{\rm bol}$ yields a radius estimate of 0.46 R$_{\odot}$.
\citet{demory2009} plot stellar radius versus absolute magnitude in the K band, M(K), for low mass and very-low mass stars using interferometric measurements; their 5 GY isochrones
together with the absolute K magnitude of the A component of the binary \citep[6.12;][]{riedel2014} suggests a radius of about 0.4 R$_{\odot}$.
These numbers are consistent with a larger radius than obtained for other nearby M dwarfs of the 
same temperature \citep[such as described in ][]{newton2015,mann2015}
and we adopt R$_{\star}$=0.4R$_{\odot}$ in this paper. 
In young stars, accretion episodes can provide an additional optical and X-ray signature to that expected from magnetic reconnection
\citep{stassun2006,brickhouse2010}. 
However, the WISE $w_{1} - w_{3}$ and $w_{1}- w_{4}$ colors of this system 
show no evidence for an infrared excess \citep{allwise}, indicating that there is no active accretion, as would be expected for stars older than 
several million years.

\begin{figure}[!h]
\includegraphics[scale=0.5]{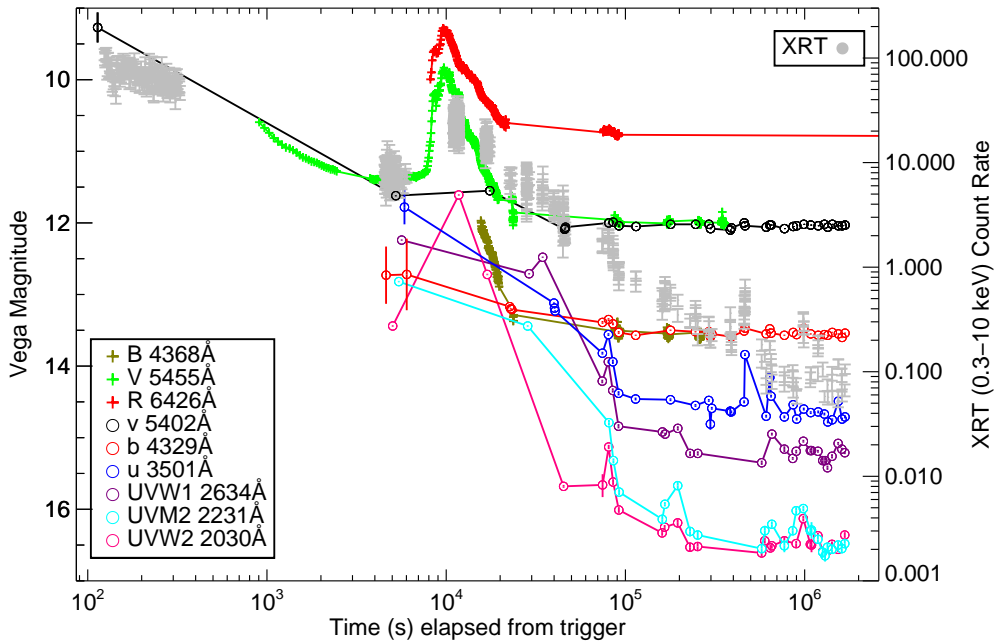}
\caption{Comprehensive light curve of the event as seen in soft X-rays, UVOT bands, and ground-based
optical photometry. The initial impulsive event took only a few hours to decay, but was followed by a series of
flares which spanned more than two weeks. The legend lists the UV optical filters and the
central wavelength of each.
\label{fig:lc}}
\end{figure}

On 2014 April 23, one of the 2 stars in this system flared to a level bright enough 
($\sim$3.4 $\times$ 10$^{-9}$ erg s$^{-1}$ cm$^{-2}$) in the 15-100 keV band that it triggered the Swift Burst Alert Telescope 
(BAT);
described in \citet{dgcvnATel}.
Two minutes later, after Swift had slewed to point in the direction of this source, the Swift X-ray Telescope (XRT) 
and the Ultraviolet Optical Telescope (UVOT) commenced observing this flare. These observations, as well as supporting ground-based optical and radio observations, continued 
(intermittently) for about 20 days and yielded a fascinating case history of this colossal event, the decay of 
which took more than two weeks in the soft X-ray band, and included a number of smaller superimposed secondary flares 
(see Fig.~\ref{fig:lc}). 
Recent papers have reported on additional data indicating radio and optical bursts from this system during this time period
\citep{fender2015,caballero2015}.
In this paper, we discuss the observations and their interpretation in light of the standard solar flare scenario.
The paper is organized as follows: \S 2 describes the entire set of Swift and ground-based observations used in the study,
\S 3 describes the analysis of the two main flaring events observed, \S 4 discusses what can be determined for the
second event, and applies this to an interpretation of the first event. Finally, \S 5 concludes.

\section{Observations }

\subsection{Swift/BAT Data}
\begin{figure}[!h]
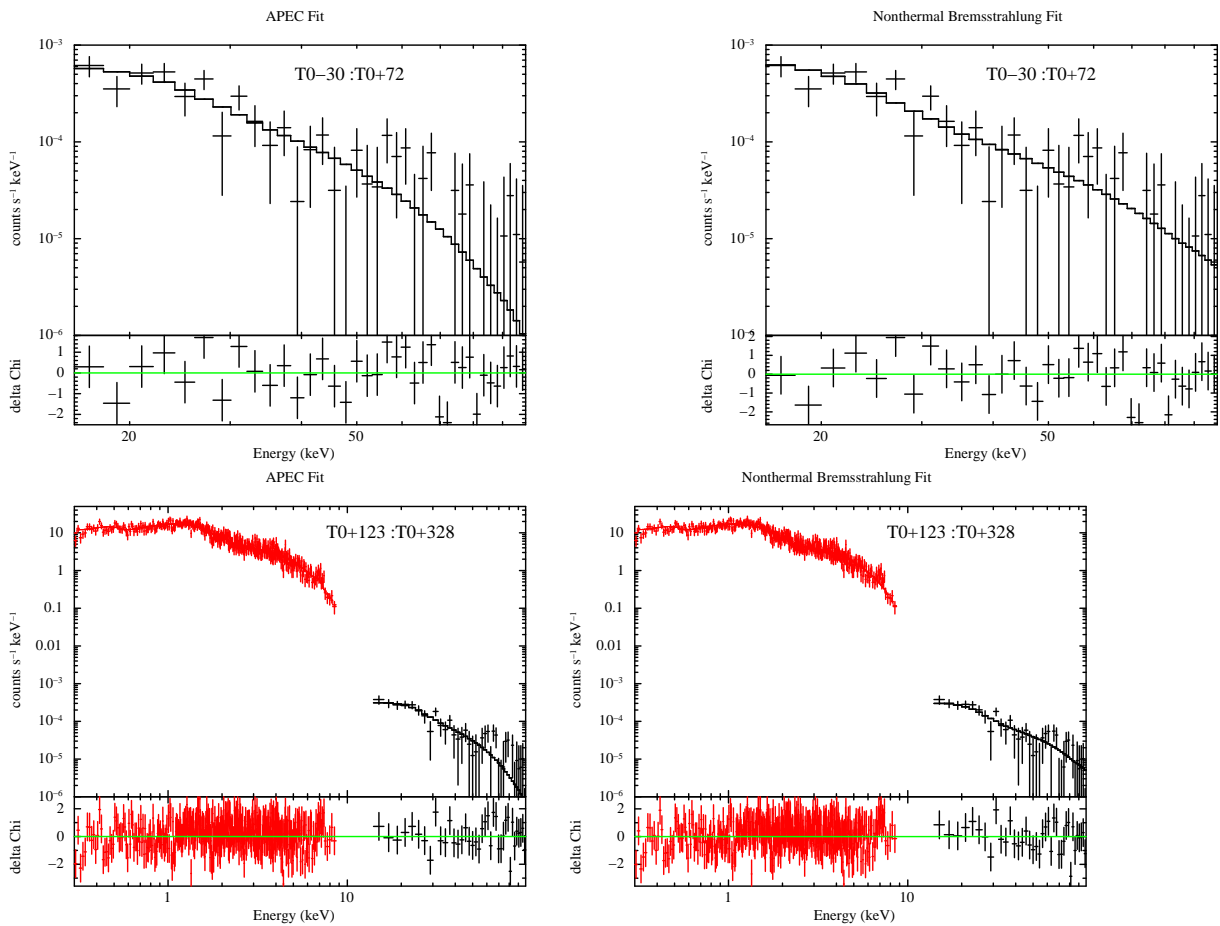

\includegraphics[scale=0.3,angle=-90]{f2a}
\includegraphics[scale=0.3,angle=-90]{f2b}
\includegraphics[scale=0.3,angle=-90]{f2c}
\includegraphics[scale=0.3,angle=-90]{f2d}
\caption{ X-ray spectra at the time of the trigger \textit{(top panels, BAT only)} and during the initial $\sim$200 seconds after the
XRT and BAT both were on source \textit{(bottom panels)}.  
The left panels show spectra fit with thermal models; right panels show same spectra fit with nonthermal models.
Spectral fit parameters are given in Table~\ref{tbl:specfit}.
\label{fig:spec}}
\end{figure}

The Swift BAT instrument \citep{batref} triggered on the flare from DG CVn at 2014 April 23 UT 21:07:08.0 = T0 (BAT
trigger number 596958).
The source location was in the BAT field of view (FOV) starting from T-1627 s, but there was no detectable emission 
until approximately T$-$40 s.  After the trigger occurred, a slew placed the star in the apertures of the narrow-field 
instruments (XRT and UVOT), for 210 seconds on target 
before an observing constraint led to a slew to another target.  However, DG CVn remained in the BAT FOV 
during this new pointing, until T+892 s.  The mask-weighted lightcurve shows a single peak from $\sim$T0-40 s to 120 s and 
another weaker peak from $\sim$T0+200 to T0+240 s. BAT spectra from 15 to 150 keV were extracted for the time intervals from 
T0-30 s to T0+72 s and from T0+123 to T0+328 s (the latter to match the initial XRT observation), and are shown in
Fig.~\ref{fig:spec}.

\subsection{Swift/XRT Data}
The XRT \citep{xrtref} began observing DG CVn 117 seconds after the
BAT trigger. 
The online XRT product
generator\footnote{\verb+http://www.swift.ac.uk/user_objects/+} \citep{evans2007,evans2009}
was utilised to produce the XRT light curve and extract the
time-sliced spectra. This tool was used to account for pile-up and to apply all
necessary corrections. At the time of writing, the software version used
by the generator was HEASoft 6.18, with the calibration file release of
2016-01-21.

For the initial snapshot of data, immediately following the BAT trigger,
the data were collected in Windowed Timing (WT) mode, due to the large
count rate. Observations between 4.5 and 50 ks after the trigger occurred
using a combination of both WT and Photon Counting (PC) modes; all later
data were then taken in PC mode.

\subsection{Swift/UVOT Data}

The UVOT \citep{roming} began observing DG CVn, 108 s after the Swift-BAT trigger, T0, with a 10s settling
exposure. After a 4.2 ks gap in XRT/UVOT observations, UVOT observed in all 7 UVOT filters with regular cadence
until 1.7Ms after the trigger. The UVOT returned to the field 4 months later (11Ms after the trigger) to determine
the quiesent level in the optical and UV filters. Upon examination of the initial images we found that all the
white exposures were saturated and therefore no futher exposures were taken later than 11.5ks. The \vfilt\ settling
image and the first two \bfilt\ and \ufilt\ exposures were also saturated. 
Observations continued from 74ks for the \vfilt, \bfilt\ and \ufilt\ filters in hardware mode
since the quiescent \Vfilt\ band magnitude for DG CVn is close to the brightness limit of the standard UVOT image mode.
In this case a smaller portion of the detector was read out,
reducing the frame time of each exposure from 11 ms to 3 ms and enabling brighter objects to be observed.

To perform the photometry for non-saturated images, we used a region of 5$^{\prime\prime}$ radius to extract the source counts
and background counts were extracted using two circular regions of radius 12$^{\prime\prime}$ from a blank area of sky situated near to the 
source position. The count rates were obtained from the images using the Swift tools uvotsource. For the saturated images
we were able to extract photometry from the first \vfilt\ exposure, the first two \bfilt\ exposures and the second \ufilt\ band exposure
using the read out streaks associated with DG CVn, taking advantage of the method recently developed by \citet{page2013}. The
resulting error bars for these exposures reflect the larger photometric uncertainty using this method. We note that the
first \vfilt\ exposure is trailed, however because this exposure was observed in event mode, we were able to extract a new
image excluding the affected, first 2.2s. The resulting count rates from both photometric methods were converted to magnitudes
using the UVOT photometric zero points \citep{breeveld2011}.

\subsection{University of Athens Observatory \Rfilt-band Data}
Relative differential photometry in optical (Bessell) \Rfilt\-band was
obtained on 23 \& 24 April 2014 with the automated and remotely
controlled 0.4 m f/8 Cassegrain telescope, equipped with an SBIG
ST10XME CCD camera and an f/6.3 focal reducer,
at the University of Athens
Observatory. 
The flare was observed from T0+8164 s to
T0+21412 s and then again from T0+75155 s to T0+91971 s. The
first 22 observations were removed due to light clouds which affected
the photometry. The cadence varied between 45 and 95 s, and relative
flux measurements were made using nearby comparison stars. Additional
observations in optical (Bessell) \Bfilt\ and \Rfilt\-bands were also obtained on
4 \& 5 August 2014, to confirm the quiescent level. Raw images
were corrected for dark current and reduced using sky-flat
images. The aperture photometry package Munipack \citep{munipack}
was used for data reduction and extraction of magnitudes and
errors. The resultant light curve is shown in Figure~\ref{fig:lc}.


\subsection{Photometry from Upice Observatory}
The remotely controlled telescope at Upice Observatory in the Czech Republic was used to observe the field
of DG~CVn with a 20 cm Newtonian telescope, and SBIG CCD camera using \Bfilt \Vfilt \Ifilt\ filters for photometry. 
Observations commenced 900 s after the initial trigger in the case of V filter observations,
and spanned 4.3 hours. 
Additional observations the following nights -- 
three additional nights in the case of the \Bfilt\ filter and about 3 weeks for the \Vfilt\ filter -- 
were also obtained to examine
long timescale variability. 
The cadence of observations in \Bfilt\ filter on the first day was about 30 s, increasing
to 80 s on subsequent days, and for the \Vfilt\ filter the corresponding cadences were about 60 s
on the first day and 126-195 s on subsequent days.
Raw images were corrected for dark current and the field was reduced using flat-field images combined from
many images obtained during various sessions to get the resultant flat background.
The aperture photometry package Munipack \citep{munipack}
was used for data reduction and extraction of magnitudes and errors.
Data are shown in Figure~\ref{fig:lc}.

\subsection{Photometry from Observatorio del CIECEM}
Data on the first night were obtained with two 14-inch Schmidt-Cassegrain telescopes
(one for \Vfilt-band observations and the other for \Bfilt-band) of the Observatorio del CIECEM in
Huelva, Spain. 
The \Vfilt\ filter observations began starting at T0+10173 s and lasted 3.8 hrs;
\Bfilt\ filter observations commenced 15610 s after the trigger, and spanned 2.3 hrs.
The observing cadence for the \Vfilt\ filter data was about 20 s on the first day, with occasional
longer cadences due to individual bad data points.
Observations proceeded over the next 5 days, spanning about 2 hrs on each of the next nights. 
Observations on these subsequent nights utilized a single 11 inch Schmidt-Cassegrain telescope, by alternating
\Vfilt\ and \Bfilt\ filters in the time series. An exposure time of 30 s was used for the \Vfilt\ band data, and 40 s for \Bfilt\ band.
Data are shown in Figure~\ref{fig:lc}.
In combining data from different telescopes using the same filter, we find calibration differences to be negligible.

\section{Data Analysis}
In addition to the multi-wavelength data presented in this paper, we also make use of radio data presented in \citet{fender2015},
and the \Vfilt\ band photometry prior to the Swift trigger presented in \citet{caballero2015}.
In this paper we concentrate on the two main events evident in Figure~\ref{fig:lc}: the big first flare (BFF), 
which appears to
extend from T-40 s based on the timing reported in \citet{caballero2015} until possibly later than T+328 when Swift ceased monitoring; and F2, whose peak is at about T0+10$^{4}$s.
Given the amplitudes, durations, and multi-wavelength data for these two events, there is much more that we can say about them.

\subsection{X-ray Spectra}
\subsubsection{Trigger and XRT+BAT spectra of BFF\label{sec:ntfits}}
We fit the spectra obtained from the time intervals T0-30:T0+72 and T0+123:T0+328 from the BAT only (first interval)
and XRT+BAT (second interval), using either a single temperature APEC (Astrophysical Plasma Emission Code) model or a nonthermal thick-target bremsstrahlung
model. 
The amount of interstellar absorption was fixed to an N$_{H}$ value of 4.7$\times$10$^{17}$ cm$^{-2}$,
based on Mg~II and Fe~II column densities measured by \citet{malamut2014} towards $\beta$~Com, a star
$\sim$4.5$^{\circ}$ away in angular extent and with a proximity of 9 pc (S. Redfield, private communication).
APEC describes a collisionally ionized plasma in coronal equilibrium, with line emission
formed predominantly from a balance of collisional 
excitations populating excited ionic states, and radiative de-excitations; this is the usual assumption for stellar
coronal plasmas.
We use a custom version of ATOMDB \citep{apecref} calculated out to 100 keV\footnote{available from 
\verb+http://www.atomdb.org/download_process.php?fname=atomdb_v2_0_2_runs+
}, since
the standard ATOMDB energy grid delivered with XSPEC ends at 50 keV.
The Volume Emission Measure ($\mathscr{VEM}$) quantifies the amount of plasma emitting at the fitted temperature,
and is equivalent to $\int n_{e}^{2} dV$.
The nonthermal thick target model is often used to describe hard X-ray emission from solar flares.
In this model, a power-law distribution of electrons with energy (described by the parameter $\delta_{X}$)
is modified by transport through a fully ionized plasma.  From the observed spectrum the index $\delta_{X}$
can be derived, as well as (for unresolved stellar observations) the power in the electron beam \citep[see 
discussion in][]{osten2007}.
In solar flare observations a broken power-law is often observed, 
with low energy cutoff ($E_{0}$) around 20 keV; 
in this case we consider a
single power-law, with $E_{0}$ fixed to be 20 keV.
The power required depends on the low energy cutoff as $(20/E_{0})^{(\delta_{X}-2)}$ as described in \citet{osten2007},
so decreasing the low energy cutoff from 20 keV to 10 keV results in roughly a factor of two more power required for
$\delta_{X} \sim$3.
The spectra are
shown in Figure~\ref{fig:spec} and fit parameters are listed in Table~\ref{tbl:specfit}.
The 0.01-100 keV flux extrapolated from the best-fit model is also reported in Table~\ref{tbl:specfit}.
The two models are statistically indistinguishable for the same time interval.

\begin{deluxetable}{llll}
\tablewidth{0pt}
\tablecolumns{3}
\tablecaption{Spectral fit parameters for trigger and initial decay of flares on DG~CVn 
\label{tbl:specfit}}
\tablehead{ \colhead{Parameter} & \colhead{T0-30:T0+72} & \colhead{T0+123:T0+328\tablenotemark{a}} \\
\colhead{} & \colhead{BAT only} & \colhead{XRT+BAT} }
\startdata
\hline
 \multicolumn{3}{c}{Thermal Fit} \\
\hline
T$_{X}$ (10$^{6}$K) & 278$^{+140}_{-92}$ & 290$\pm31$ \\
$\mathscr{VEM}$ (10$^{54}$ cm$^{-3}$) & 9$^{+4.7}_{-2.7}$  & 4.9$\pm$0.1 \\
$\chi^{2}$ (dof) & 38.2 (36) & 401.6 (432) \\
\multirow{2}{2in}{f$_{X}$ (14-100 keV) $\times$10$^{-9}$ \\(erg cm$^{-2}$ s$^{-1}$)}  & 3.6$^{+0.2}_{-0.8}$ & 2.03$^{+0.18}_{-0.15}$  \\
    &                     &              \\
\multirow{2}{2in}{f$_{X}$ (0.3-10 keV) $\times$10$^{-9}$ \\ (erg cm$^{-2}$ s$^{-1}$)}  & (4.5)\tablenotemark{1} & 2.38$\pm0.03$  \\
    &                     &              \\
\multirow{2}{2in}{f$_{X}$ (0.01-100 keV) $\times$10$^{-9}$ \\ (erg cm$^{-2}$ s$^{-1}$)} & 9.3 & 5.1 \\
    &                     &              \\
\hline
\multicolumn{3}{c}{Nonthermal Fit} \\
\hline
$\delta_{X}$ & 3.6$\pm$0.4 & 3.2$^{+0.2}_{-0.1}$ \\
Power (10$^{37}$ erg s$^{-1}$) & 3$^{+1.2}_{-1.1}$  & 1.11$\pm0.03$  \\
$\chi^{2}$ (dof) & 40.2 (36) & 409.62 (432) \\
\multirow{2}{2in}{f$_{X}$ (14-100 keV) $\times$10$^{-9}$ \\ (erg cm$^{-2}$ s$^{-1}$)} & 3.8$^{+0.2}_{-0.4}$ & 2.24$^{+0.19}_{-0.15}$\\
	    &                     &              \\
\multirow{2}{2in}{f$_{X}$ (0.3-10 keV) $\times$10$^{-9}$ \\ (erg cm$^{-2}$ s$^{-1}$)}  & (5.8)\tablenotemark{1} & 2.34$^{+0.03}_{-0.02}$ \\
    &                     &              \\
\multirow{2}{2in}{f$_{X}$ (0.01-100 keV) $\times$10$^{-9}$ \\ (erg cm$^{-2}$ s$^{-1}$)} & 11. & 5.3 \\
    &                     &              \\
\enddata
\tablenotetext{a}{N$_{H}$ fixed at 4.7$\times$10$^{17}$ cm$^{-2}$.}
\tablenotetext{1}{Flux extrapolated from best-fit model in the 14-100 keV range.}
\end{deluxetable}

The thermal fit shows that the spectrum from 0.3-100  keV for each time interval is dominated by a single temperature.  Attempts to use a more complicated model, 
like multiple temperature components, did not result in a statistically better fit,
demonstrating the dominance of the hote temperature plasma.
We confirmed that the spectrum 
is dominated by the continuum from such a hot plasma by redoing the fit using a bremsstrahlung model (brems) in 
XSPEC, and comparing the result to the APEC fit.  The bremsstrahlung model only includes contributions from H and He,
which confirms the high temperature results from fitting with APEC.  There is no evidence for any lower temperature 
plasma, as evidenced by the lack of He- or H-like iron at characteristic energies of 6.7 and 6.9 keV.  

Previous reports of stellar superflares with Swift had reported the detection of the Fe K$\alpha$ line at 6.4 keV
\citep{osten2007,osten2010}. However, these appear to be due to a calibration artifact unrecognized at the time, namely
charge trapping \citep{pagani2011}. The spectral fits to XRT data shown in Figures~\ref{fig:spec} for BFF and
~\ref{fig:f2fit} for F2 do not show any evidence of excess emission near 6.4 keV. 



\subsubsection{XRT spectra of F2 \label{sec:f2}}
\begin{figure}[!h]
\includegraphics[scale=0.6,angle=-90]{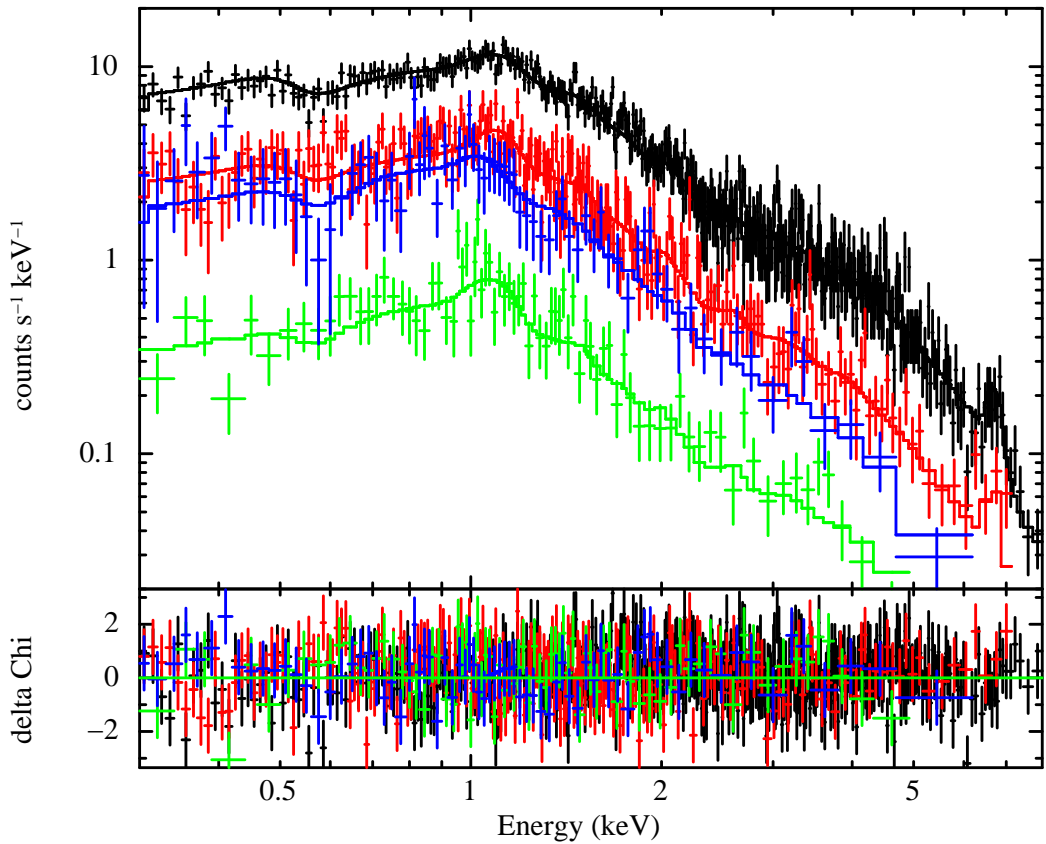}
\caption{Top panel shows four spectra from the F2 event: 
black is T0+10860:T0+11849, red is T0+16496:T0+17605,
green is T0+22437:T0+23367, and blue is T0+28019:T0+29125.
Solid histogram gives the best-fit spectral model for that
time interval; parameters are listed in Table~\ref{tbl:f2specfit}. Bottom panel gives the contribution to 
the $\chi^{2}$ statistic in each energy bin, for each dataset and model. 
\label{fig:f2fit}
}
\end{figure}

We extracted XRT spectra in four intervals during the peak and decay of the second large flare, F2. There was not enough
signal in the BAT at this time to extract a spectrum, so we concentrate on the 0.3-10 keV range of the XRT.  
The four spectra were fit jointly, with a three temperature APEC model: the two lowest temperatures (corresponding to
quiescent emission) were fixed to be the same for all four spectra, while the third temperature component
corresponding to the flare emission
was allowed to vary. 
We arrived at this after comparing the goodness-of-fit for models with different numbers of components.
Values are given in Table~\ref{tbl:f2specfit}. Columns labelled ``Q1" and ``Q2" list the results for the 
quiescent component; the metallicity was fixed to be unity (i.e. solar) for the spectral components. 
Spectral fits extracted for other times during the peak and decay of the flare are listed in separate
columns. The temporal trends show a large temperature and volume emission measure at the earliest time interval,
with generally decreasing plasma temperatures and volume emission measures during the decay of the flare.
The abundance of the flaring plasma initially appears to be mildly sub-solar, but noise in the values
fitted from spectra extracted
at later times prevents determining a definitive trend.
Since the spectra were fit together, a single
value of the fit statistic was calculated: $\chi^{2}$ of 749.81, with 787 degrees of freedom.
Figure ~\ref{fig:f2fit} shows the spectra along with the model fits. Table~\ref{tbl:f2specfit} also gives a flux
extrapolated to the 0.01-100 keV range for intercomparison of the different time intervals and other flares.

\begin{deluxetable}{lllllll}
\tablewidth{0pt}
\tablecolumns{7}
\rotate
\tabletypesize{\scriptsize}
\tablecaption{Spectral fit parameters during F2 event on DG~CVn \label{tbl:f2specfit}}
\tablehead{ \colhead{Parameter} &\colhead{Q1} & \colhead{Q2} & \colhead{T0+10860:T0+11849} & \colhead{T0+16496:T0+17605}
&\colhead{T0+22437:T0+23367} & \colhead{T0+28019:T0+29125} }
\startdata
T$_{X}$ ($10^{6}$K) &  5.0$^{+2.4}_{-1.3}$ & 14.7$^{+2.7}_{-2.8}$  & 53.8$^{+2.9}_{-3.8}$ & 41.5$^{+4.9}_{-4.4}$  & 
55.7$^{+20.2}_{-13.0}$  & 36.2$^{+11.3}_{-7.8}$  \\
$\mathscr{VEM}$ (10$^{52}$ cm$^{-3}$) & 1.3$_{-0.6}^{+0.8}$ &3.6$^{+2.3}_{-2.0}$ & 154.9$^{+5.5}_{-6.3}$ & 105.3$_{-9.0}^{+8.6}$ & 38.6$^{+8.3}_{-11.2}$ & 43.0$^{+8.2}_{-8.6}$ \\
Z & 1 (fixed) & 1 (fixed) & 0.4$_{-0.12}^{+0.12}$ & 0.51$^{+0.24}_{-0.20}$ & 1.18$^{+2.12}_{-0.88}$ & 0.05$^{+0.52}_{-0.05}$ \\
\multirow{2}{2in}{f$_{X}$\tablenotemark{1} (0.3-10 keV) $\times$10$^{-10}$\\ (erg cm$^{-2}$ s$^{-1}$)}& \textit{0.09}\tablenotemark{2}&\textit{0.20}\tablenotemark{2} & 
6.59$^{+0.07}_{-0.11}$ & 4.35$^{+0.11}_{-0.11}$ & 2.26$^{+0.10}_{-0.19}$ & 1.60$^{+0.14}_{-0.06}$ \\
	& & & & & &  \\
\multirow{2}{2in}{f$_{X}$\tablenotemark{3} (0.01-100 keV) $\times$10$^{-10}$  \\ (erg cm$^{-2}$ s$^{-1}$) }& \textit{0.13} &\textit{0.28} & \textit{8.0} & \textit{5.3} & 
\textit{2.8} & \textit{2.0} \\
	& & & & & &  \\
\enddata
\tablenotetext{1}{Flux calculated for flaring time intervals includes contribution from quiescent plasma component.}
\tablenotetext{2}{Quiescent flux in this energy range calculated using best-fit model parameters.}
\tablenotetext{3}{Flux extrapolated into this energy range using best-fit model parameters.}
\end{deluxetable}

\subsection{ Flaring Footpoint Emission \label{sec:foot}}
We use the ground-based \Vfilt\ band, \Rfilt\ band, and Swift UVOT data to determine the white-light emitting footpoint sizes during the peak of BFF and F2.
We use the following equation to determine the fraction of the visible hemisphere, $X_{\rm flare}=R_{\rm flare}^2/R_{\star}^2$, producing white-light
emission \citep{Hawley2003}
\begin{equation} 
F_{\rm{flare, filter}} = I_{\rm f,filter} \times F_{\rm{Q}, \rm{filter}} = X_{\rm{flare,filter}} \left( R_{\star}^2 / d^2\right) \pi B_{\lambda=\lambda_{eff}}(T_{\rm{flare}}) 
\end{equation}
where $\pi R_{\rm flare}^{2}$ is the total flaring area, F$_{\rm{flare, filter}}$ is the observed peak flare flux in a particular filter,
$I_{\rm f, filter}$ is the fractional increase relative to the quiescent flux during the flare, $F_{\rm{Q}, \rm{filter}}$ is the quiescent flux in that filter,
$B_{\lambda=\lambda_{\rm eff}}(T_{\rm{flare}})$ is the Planck function evaluated at the effective wavelength $\lambda_{\rm eff}$ of the filter,
and $T_{\rm flare}$ is the temperature of the blackbody flare spectrum assumed to dominate the spectral energy distribution of the optical filter.
The quantities $R_{\star}$ and $d$ are the radius of the star and its distance from Earth, respectively.
We use this equation initially to calculate $X_{\rm flare,filter}$ for the peaks of both BFF and F2 assuming 
a $T=10^4$ K blackbody flare spectrum, which is a reasonable approximation to the peak broadband color distribution during very large flares \citep{Hawley1992}.
\citet{riedel2014} list the \Vfilt-band magnitude of the DG~CVn system as \Vfilt=12.02, which we take as the quiescent value
per star.
We assume both stars contribute equally to the brightness, so the quiescent \Vfilt-band flux is
2.87$\times$10$^{-14}$ erg cm$^{-2}$ s$^{-1}$ $\AA^{-1}$.
We take the quiescent flux to be 5.4$\times10^{-14}$ erg cm$^{-2}$ s$^{-1}$ \AA$^{-1}$ for the \Rfilt-band and 2.9$\times10^{-14}$ erg cm$^{-2}$ s$^{-1}$ \AA$^{-1}$ for the
Swift v.
We use the average of \uvwfilt\ fluxes at t-T0$>$1.3$\times$10$^{6}$ s to determine the quiescent flux for \uvwfilt\.
All values of $I_{f}$ were calculated relative to one star, assuming the quiescent flux from the system can
be divided equally between the two components).
The peak \Vfilt-band magnitude increase for BFF is $\Delta \Vfilt=-$5 mag from \citet{caballero2015}, so the relative flux increase for one star 
is $I_{f, V}$ $\sim$190, whereas the v-band from Swift in the decay phase was at $I_{f,v}$ =23.5 at $T0+113.5$~s.
For the Swift \vfilt\ measurement of BFF, there is an 
uncertainty of 0.21 mag due to saturation and spillover, so the measurement and uncertainty in $X_{\rm{flare,filter}}$
give $X_{\rm{BFF,v}} \sim 0.044$ ($0.035 - 0.054$).
The first v-band point from Swift is consistent with the \Vfilt-band photometry from \citet{caballero2015}
in the decay phase of BFF.
For F2, $I_{f,V}=$11.8 at the flare peak, and I$_{f,uvw2}=$156. 
These lead to 
$X_{BFF,V} (10^{4}K)$ = 0.375 for BFF and 
$X_{F2,V} (10^{4}K)$ = 0.023 for F2 based on \Vfilt-band measurements.
In the \Rfilt-band, $I_{f,R}=5.9$ at the peak of F2 ($T0+9590$~s), leading to a value $X_{\rm F2,R} (10^{4}K)$= 0.031.
These values are summarized in Table~\ref{tbl:footpoint}.
These footpoint sizes are extraordinary for M dwarf flares and are comparable to those inferred during the flares from \citet{osten2010} and \citet{Schmidt2014}, but 
10 times larger than the peak footpoint sizes in the Great Flare on AD Leo \citep[][hereafter, HP91]{hawleypettersen}.

\begin{deluxetable}{lllll}
\tablewidth{0pt}
\tablenum{3}
\tablecolumns{5}
\tablecaption{Implied Flaring Footpoint Area Fractions \label{tbl:footpoint}}
\tablehead{ 
\colhead{Time} &\colhead{filter} & \colhead{I$_{f}$} & \colhead{T$_{BB}$} & \colhead{X$_{\rm flare}$}\\
\colhead{(s)} & \colhead{} & \colhead{} & \colhead{(K)} & \colhead{}
 }
\startdata
\multicolumn{5}{c}{BFF}\\
T0-40 & \Vfilt & 190 & 10$^{4}$ & 0.375 \\
T0+113.5 & \vfilt\ & 23.5 & 10$^{4}$ & 0.044 ($0.035 - 0.054$)\\
\multicolumn{5}{c}{F2}\\
T0+9590 & \Rfilt\ & 5.9 & 10$^{4}$ & 0.031 \\
T0+9590 & \Rfilt\ & 5.9 & 6000 &  0.15 \\
T0+9710      & \Vfilt &  11.8  & 10$^{4}$ & 0.023 \\
T0+9710      & \Vfilt & 11.8   & 6000 &  0.14\\
T0+ 11761     & \uvwfilt\    &  156  &10$^{4}$ &  0.005    \\
T0+ 11761     & \uvwfilt\    &  156 &6000 &   0.59  \\
T0+11753  & \Vfilt & 5.1 & 10$^{4}$  & 0.01\\
T0+11753  & \Vfilt & 5.1 & 6000  & 0.064\\
\enddata
\end{deluxetable}



While a T$=$10$^{4}$K blackbody is reasonable for the impulsive phase of very large flares,
the flare (F2) exhibits much longer timescales than BFF and may also exhibit different heating properties
(e.g. through the apearance of a Vega-like spectrum; K13).
Thus, we search for evidence of a different color temperature using the \Vfilt\ and \Rfilt\ band data during F2.
The right panel of Figure~\ref{fig:Renergy} displays the \Vfilt/\Rfilt\ ratio during the decay phase of F2 in addition to the detailed \Vfilt\ and \Rfilt\ light curves,
and reveals a ratio with a maximum value near 1.2 at the time of the peak of F2.
A \Vfilt/\Rfilt\ ratio of $\sim$1 was also observed in the Great Flare on AD~Leo
\footnote{\Rfilt\ in this flare was obtained in Johnson \Rfilt\ which is redder and wider than Bessell R} in the gradual phase (HP91) 
and a \Vfilt/\Rfilt\ ratio $\sim$1 was synthesized from spectra during the decay of the Megaflare on YZ CMi
\citep{Kowalski2010}.
The \Vfilt/\Rfilt\ ratio indicates a lower blackbody temperature, of about 6000 K, for F2, and we also calculate the X$_{\rm flare}$ values
for this temperature as well in Table~\ref{tbl:footpoint}.
Additional contributions from Balmer continuum emission are expected in the Swift UVOT filter bandpasses
\citep[see discussion in][]{Kowalski2010} 
and a redder continuum component, termed ``Conundruum'' in \citet{Kowalski2013} (hereafter K13), is also expected in the \Rfilt-band.
The
 broadband UVOT and \Rfilt-band distribution will be discussed in more detail in future work 
(Kowalski et al. 2016, in preparation).
The \Vfilt/\Rfilt\ color declines to less than 1 in the decay of the F2 event, which indicates a 
``cooler'' Conundruum component in F2 than in the Megaflare 
(closer to the color temperature of the Conundruum in the 
IF3 event from K13; Figure 31 of that paper).  Without spectra, we cannot assess the detailed properties of the changes in the 
blue continuum for the newly formed flare emission and thus we do not know the area of the newly formed kernels.  
However, the \Vfilt-band flux experiences a much larger relative increase (7x compared to the previously decaying emission 
from the BFF at T0+$\sim$5e3s in Fig~\ref{fig:Renergy}) than the synthesized \Vfilt-flux increase (~1.5x) in the 
secondary events following the Megaflare event.  
Therefore, we would expect that persistent hot spots with very blue spectra \citep[either a spectrum like a blackbody with 
T$\sim$10,000 K or a Vega-like spectrum as found in][]{Kowalski2010} would have resulted in a much larger change in the \Vfilt/\Rfilt\ flux ratio for F2.  
If we assume a value of the \Rfilt-band flux before F2 (such that \Vfilt/\Rfilt\ just before F2 is the same as the \Vfilt/\Rfilt\ in the decay at 
T0+17,000 s), we estimate that the color temperature of the newly formed emission ranges between 6000-8000 K, but 
nowhere near the \Vfilt/\Rfilt\ value (1.7) for a Vega-like spectrum.  The \Vfilt/\Rfilt\ color temperature of about 6000 K is evidence that 
increased Conundruum continuum emission dominates the optical brightness increase in the F2 event.

The Swift \uvwfilt\ point which falls near the peak of F2, but shortly afterward, can place some constraints
on the relative contribution of blackbody versus Balmer continuum expected at these short wavelengths.
No extinction correction has been applied here, but due to the proximity of the star and its location out
of the Galactic plane this will be negligible.
At T0 + 11753 to T0+ 11761, the flux ratio of \uvwfilt/\Vfilt\ is 1.2$\times$10$^{-13}$ / 1.5$\times$10$^{-13}$ = 0.8. 
At T0 + 16951, the ratio of \uvwfilt/\Vfilt\ is 4.2$\times$10$^{-13}$ / 4.6$\times$10$^{-13}$ = 0.9.  
It is interesting that X$_{F2,uvw2}$(T=6000 K) is 0.59 for \uvwfilt\ and X$_{F2,V}$=0.06 for \Vfilt\ at T0 + 11753 (see Table~\ref{tbl:footpoint}).
This indicates that 10 times more emission is needed to account for the flux in the \uvwfilt\
bandpass if a 6000 K blackbody (or any spectrum that is similar shape to a 6000 K blackbody) 
is extrapolated to $\lambda=$2030 \AA.
In the impulsive beam heating phase of the F11, F12, and F13 models from \citet{kowalski2015}, 
the average values of the continuum flux ratio 2030/5500 \AA\ are 1.8, 2.0, and 3.1 for these models respectively.  
So a ratio of 0.8 - 0.9 (even given a ~20\% uncertainty from comparing satellite and ground-based broadband data)  
can be used as a strong constraint on heating models. 
If the 6000 K blackbody is a good approximation to the \Vfilt\ and \Rfilt\ continuum in the F2 event, then there is Balmer continuum necessary to 
account for the \uvwfilt\ data point, but not as much Balmer continuum as the impulsive phase F11 or F12 models predict.


The Megaflare described by K13 was similar to the situation presented here in that a second large
flare occurred during the decay phase of a large,
 $\Delta \Ufilt=-$6 magnitude, flare brightening on the nearby flare star YZ~CMi. While the YZ~CMi events did not have
coverage by high energy satellites, they did have comprehensive blue-optical spectrophotometric coverage, which enabled
several inferences to be drawn about the behavior of the lower stellar atmosphere.
Spectra covering the green and red continuum  ($\lambda>$ 5000 \AA) in the Megaflare 
indicate the presence of a significant red continuum component  with T$_{BB}$ $\approx$ 5500 K 
(Figure 31 of K13). 
At the same time, the blue continuum ($\lambda$=4000-4800 \AA) exhibited a hotter 
($\sim$8000-8500 K) color temperature than in the red.  During the secondary flares in the decay phase of the Megaflare, the 
blue continuum increased in color temperature to 11,000 K (Table 9, K13), which was attributed to newly formed flare emission
 resembling the spectrum of Vega.  Over the secondary flares, the \Vfilt/\Rfilt\ ratio changed from 1.0 to 1.1 indicating 
that the broad continuum covering the \Vfilt\ and \Rfilt\ bands remained relatively flat, due to the brightness of the decay 
emission that was dominated by the Conundruum continuum component.  The appearance of a very blue spectrum 
(T$_{BB}$ $\sim$ 15,000 K; Table 9 of K13) in these secondary flares had a small effect on the \Vfilt/\Rfilt\ value of the total (decay + secondary flare) emission.

Invoking the solar analogy of a two ribbon flare which was applied to the Megaflare \citep{Kowalski2012}, 
the relevant flaring areas for the DG CVn superflare are the following:\\
\begin{enumerate}
\item The area of the decaying emission before F2 occurred.  This area could be decaying ribbon emission from the BFF event, 
and is likely dominated by Conundruum continuum emission and Balmer continuum emission 
(where the area fraction occupied by these two components are roughly equal).
Or it could be decaying emission from loops ignited at earlier times in F2.

\item The area of newly formed flare emission during the F2 event. This area is white-light kernel emission 
at the footpoints of newly reconnected flare loops.   This may be an area of hot blackbody-like emitting kernels 
with $X_{F2}$ (T$\sim$10,000 K), but we do not have flux-calibrated blue spectra to characterize the color temperature from 
$\lambda$=4000-4800 \AA.
\end{enumerate}

It is possible that the lifetime of the hot blackbody emission in each newly formed flare kernel is much less than the F2 
event duration of $\sim$ 3600 s and thus the blue spectra from each burst decay quickly leaving bright 
Conundruum emission to dominate the flare energy over a long timescale.  To obtain a flare footpoint area, we 
assume the following: 1) the area of the newly formed flare emission is approximately equal to the area of the Conundruum 
continuum emission (however, the loops producing Conundruum and loops producing hot blackbody emission may have different 
heating sources);  2) the Conundruum emission can be approximated by blackbody radiation (the emissivity process(es) that 
give rise to the Conundruum emission are not yet well understood; see K13); 3) the H$\alpha$, He I 5876 and Na I D line 
contributions to the \Rfilt-band are small relative to the continuum (at most 10\% is attributed to H$\alpha$ in the gradual 
phase of other large flares; HP1991 and K13).  Using T$\sim$6000 K, 
for the peak of F2, we use the \Vfilt/\Rfilt\ band ratio to determine that $X_{\rm flare}$=$X_{\rm Conundruum} =$ 0.14
(see Table~\ref{tbl:footpoint}).
The density of the plasma producing the Conundruum should be investigated with radiative-hydrodynamic ``multithread" models.


\subsection{Energetics }
\subsubsection{X-ray radiated energy\label{sec:Xengy}}
Calculation of the X-ray radiated energy is made difficult by data gaps arising from satellite occultations,
as well as limitations of data from the BAT due to signal-to-noise constraints. 
We attempt to account for the energetics in a couple of ways, to estimate the total energy for BFF and F2.
Table~\ref{tbl:specfit} lists the fluxes in two intervals of time for which flux measurements in the
14-100 keV bandpass can be made; for one of these time intervals the flux in the 0.3-10 keV interval can 
be measured directly, and in the other it can be estimated by extrapolating the best-fit spectral model into
this energy region. 
Because of the data gaps we could not do direct integration 
under the X-ray light curve. 
Instead, we assumed continuity of the flares across the data gaps 
and used exponential rises and decays to 
parameterize the light curve as a series of flare events; the parameters were not fit to the light curve, but rather
were varied to approximate the shape of the light curve. 
Figure~\ref{fig:XRenergy} shows the XRT light curve with these fits.
Thus the energy estimate done this way is only approximate.
We note that the radio 
light curve in \citet{fender2015} shows the radio flare decay of BFF in the Swift data gap, 
showing its decay to be relatively simple.  
We used energy conversion factors from the spectral modelling described above to 
determine the integrated energy in the 0.3-10 keV bandpass. 
This gives an estimated X-ray radiated energy for BFF in the 0.3-10 keV range, from T0+123 to approximately T0+10$^{4}$s,
of 4$\times$10$^{35}$ erg. We can add in the radiated energy in the 14-100 keV bandpass from T0-30:T0+328, using the
fluxes derived from spectral modelling in Table~\ref{tbl:specfit}, and the 0.3-10 keV flux in the T0-30:T0+72
time range listed in Table~\ref{tbl:specfit} and extrapolated from the best-fit model to the 14-100 keV energy range.
These last two contributions are $\approx$5$\times$10$^{34}$ erg, putting a lower limit on the X-ray radiated energy
in the 0.01-100 keV range for BFF of $\approx$4.5$\times$10$^{35}$ erg. 
We also calculate an upper limit to the energy radiated in the 0.01-100 keV bandpass
by using the ratio of flux in the 0.01-100 keV energy range to that in the 0.3-10 keV range, as listed
in Table~\ref{tbl:specfit} for BFF.
For BFF, the ratio for both time intervals in Table~\ref{tbl:specfit} is  $\sim$2, so the radiated energy
in the 0.01-100 keV bandpass could be as much as twice that deduced from the 0.3-10 keV energy range, 
or 8$\times$10$^{35}$ erg.
We take the duration of BFF to be the time from the
optical rise preceding the trigger, occuring around T0-70 s as discussed in \citet{caballero2015}, 
until the transition from BFF to F2 in the optical light curves,
at around T0+7750 s.

The F2 event is considerably longer-lasting than BFF, extending from about T0+6800 to about T0+30800 s. The decay
could be even longer; 
because
of the decrease in count rate and shorter monitoring intervals it is difficult to see whether the behavior
from about 2-5$\times$10$^{4}$ s is a continuing decay from the peak of F2 or whether there are subsequent smaller flares
occurring.  The spectral fitting for this event, in Section~\ref{sec:f2}, does suggest that the plasma is cooling.
The radiated energy in the 0.3-10 keV band for the F2 event is 9$\times 10^{35}$ erg, more than twice 
the already large energy of BFF; this is due primarily to the extended duration of F2, as its count rate
is lower. 
Calculating an upper limit for the energy radiated in the 0.01-100 keV bandpass, 
using the ratio of flux in the 0.01-100 keV energy range to the 0.3-10 keV range, we
find that 
the ratio for F2 peaks at 1.2 for the first time interval in the flare, and decreases thereafter. 
This implies
a correction of at most 20\% in the radiated energy determined for the 0.3-10 keV bandpass, or
a 0.01-100 keV radiated energy of 10$^{36}$ erg.
We take the duration of F2 to be approximately the interval T0+6800:T0+30800 s.

The radio data also reveal the presence of a radio flare at $\sim$1 day
which falls in a gap of the X-ray data. We accounted for this ``missing'' flare using an approximate
rise and decay that would fit within the gap in the X-ray data (named F5).
Accounting for the several smaller events that occurred afterwards, we estimate the total radiated X-ray energy
for the series of events spanning $\sim$19 days to be about 2$\times$10$^{36}$ erg.
We note that this is only an approximation, due to the data gaps, but suggests that F2 was responsible for 
about half of the energy release during this extended period, with BFF responsible for half again as much X-ray
radiated energy.

\begin{figure}
\includegraphics[scale=0.7]{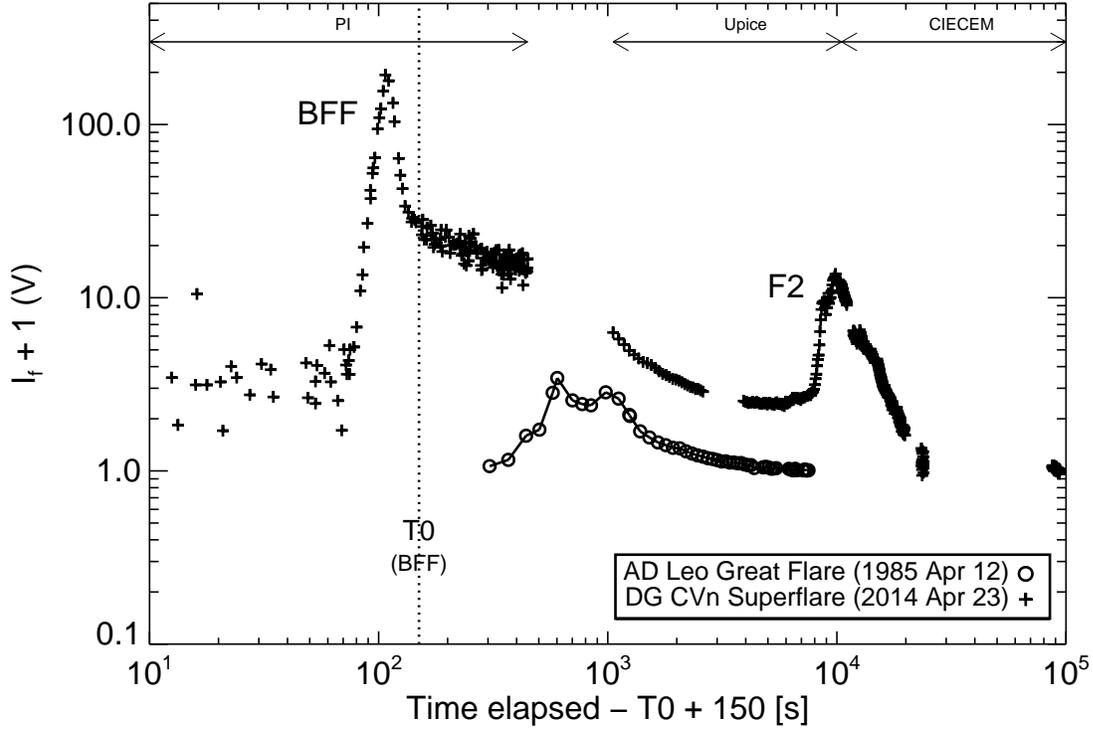}
\includegraphics[scale=0.7]{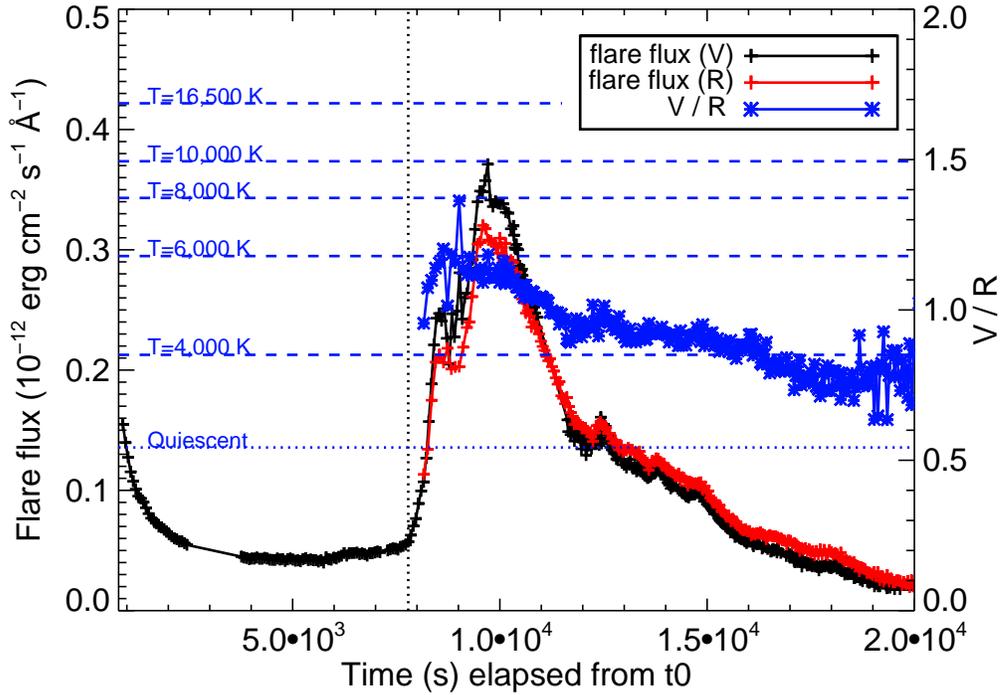}
\caption{\textbf{(top)} 
Plot of fractional increase relative to quiescent flux in the \Vfilt\ band 
for DG~CVn showing both BFF and F2 events, 
along with AD Leo Great Flare described in 
HP91 for comparison.  
The disparity in timescales between BFF, with a FHWM of 20 s, and F2, with a FWHM of 3600 s, is apparent.
\textbf{(bottom)} Comparison of \Rfilt\ and \Vfilt\ band light curve of F2 event, along with flux ratio.  
Blue dashed lines indicate approximate value of \Vfilt/\Rfilt\ ratio expected for a flare spectral energy distribution
dominated by a blackbody with the temperatures listed.
\label{fig:Renergy}
}
\end{figure}

\begin{figure}[h]
\includegraphics[angle=90,scale=0.5]{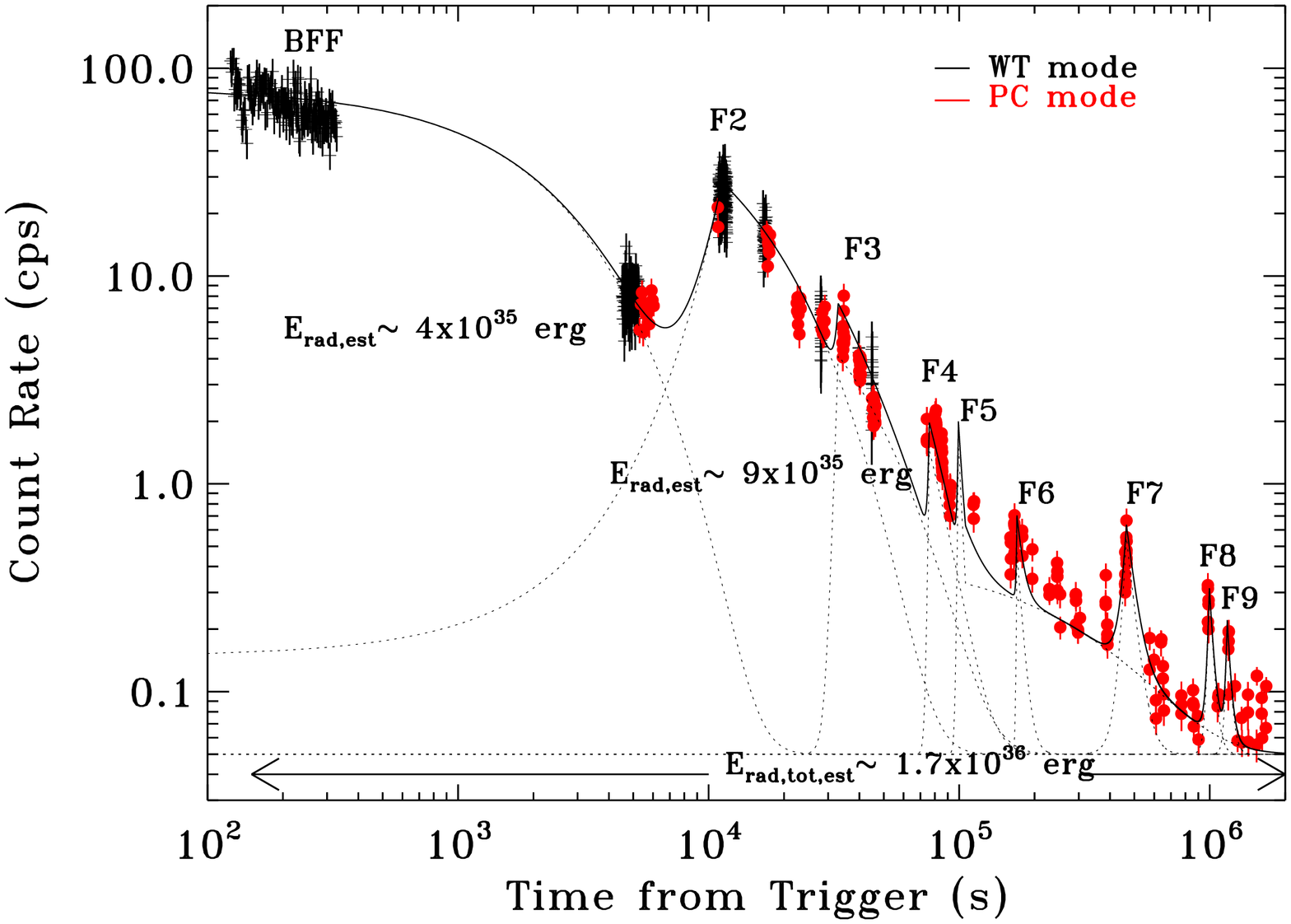}
\caption{Swift XRT light curve of the coverage of the initial trigger and subsequent flaring activity of DG~CVn
over the following $\sim$21 days.  Because of gaps in the data, exponential rises and decays were used to approximate
the observed behavior, and the models were integrated in time to estimate flare radiated energies in the 0.3-10 keV
bandpass.
\label{fig:XRenergy}}
\end{figure}

\clearpage

\subsubsection{\Rfilt\ Band Radiated Energy\label{sec:Rengy}}
The \Rfilt-band light curve of F2 peaks at T0+9590s and has a duration of 3.62 hours (from T0+8165 sec 
to T0+21,200 sec).  At the beginning of the \Rfilt-band observations, DG CVn was already 0.78 mag brighter 
than quiescence, and was increasing in brightness.  The observed duration of the rise phase is 1300 sec, and 
at the peak of F2, the system is 1.49 mag brighter than quiescence.  The flare energy is the quiescent luminosity 
multiplied by the equivalent duration of the flare \citep{Gershberg1972}.  
The equivalent duration (of one star) for F2 is 2.84$\times10^{4}$ sec, giving an \Rfilt-band flare energy of $E_{\rm{F2}, R} = 8.5 \times 10^{34}$ erg.
Figure~\ref{fig:Renergy} shows the \Vfilt\ band flare light curve, along with that of the ``great flare'' of AD Leo from 
HP91 for perspective.

\subsubsection{\Vfilt\ Band Radiated Energy\label{sec:Vengy}}

\Vfilt\ band light curves are available for both BFF and F2: the former from \citet{caballero2015}, and the
latter in this paper.
From inspection of Figure~1 of \citet{caballero2015}, it is clear that the system was slightly elevated at the start
of their optical data, before the burst beginning around t-T0=-70 s which corresponds to BFF.
The disparity in timescales in the \Vfilt\ band between BFF and F2 is apparent in Figure~\ref{fig:Renergy},
with a FWHM duration of 20 s for BFF, and $\sim$3600 s for F2.
We translated the light curve data from
\citet{caballero2015}'s Figure 1, using GraphClick, 
to estimate what the integrated \Vfilt\ band energy of BFF might be, in combination
with \Vfilt\ band measurements presented here which reveal the light curve behavior just prior to F2.
This corresponds to an integrated energy in the \Vfilt\ band of
about 2.8$\times$10$^{34}$ erg for BFF, from T0-146 to T0+7730.

We estimate the \Vfilt\ band radiated energy for F2 using direct integration from the light curve. 
From T0+7730 to T0+23754, the 
integrated energy is 5.2$\times$10$^{34}$ erg.  The light curve behavior in Figure~\ref{fig:lc} for the \Vfilt\ band
appears to demonstrate that the system has returned to its quiescent value after F2, although the gaps in the 
photometry prevent a definitive statement.  
The impulsive phase initiation of F2 is obvious from the light curve, although the decay from 
BFF flattened out for a long time before this, being $\sim$ 0.6 magnitudes above the quiescent system
\Vfilt\ magnitude. The fact that the \Vfilt\ band photometry was also enhanced prior to the BFF burst at $t-T0=-70$ s
suggests that there may have been an even earlier event missed in all wavelengths.

\subsubsection{Energy Partition in BFF and F2 and Estimates of Total Radiated Energy \label{sec:epart}}
Table~\ref{tbl:epart} lists the X-ray and optical energies derived for both BFF and F2, as discussed
in Sections~\ref{sec:Xengy}, ~\ref{sec:Rengy}, and ~\ref{sec:Vengy}. We have measurements of the X-ray and
\Vfilt\ band energy for both events, as well as (for F2) radiated energy in the \Rfilt\ band. 
F2 is the more energetic of the two events; there also appears to be a difference
of about a factor of two in the energy of the two events in both the 0.3-10 keV X-ray band and \Vfilt-band radiated energies.
We estimate the \Ufilt\ band energy using E$_{U}$-E$_{V}$ scalings from \citet{hawleypettersen}. 
Although the X-ray radiated energy in the 0.3-10 keV band appears to differ only by about a factor of two for
the two events, the high plasma temperatures derived for BFF increase the wavelength range
over which significant emission is received. As discussed in \S~\ref{sec:Xengy}, we can come up with a 
range of the likely coronal radiated energy by considering a wider photon energy range. 
Consideration of this wider wavelength range increases the amount of radiated energy from coronal plasma,
up to the point where the coronal radiated energy from BFF is only slightly less than that from F2.
This suggests that the energy partition in each event does not follow the same trend.

\citet{ostenwolk2015} described energy partition in solar and stellar flares using smaller stellar flares,
and demonstrated a rough agreement between solar and stellar flares in the relative fraction of radiated
energy appearing coronal plasma and that in the hot blackbody emission which dominates the \Ufilt-band.
From their Table~2 we estimate the bolometric radiated energy two ways: using the extrapolated \Ufilt\ band
energy described above, as well as the X-ray energies calculated in \S~\ref{sec:Xengy}.
The estimation of bolometric energy using the coronal radiated energy is imprecise because the formulation
of \citet{ostenwolk2015} only considered the 0.01-10 keV energy range, whereas the BFF event clearly
has significant contribution at higher photon energies. The two estimates of bolometric energy differ from
each other by a factor of 3 or more.  If the same relative contribution does hold for all of the coronal
plasma, then accounting for the upper limit to the total coronal radiated energy in the 0.01-100 keV 
for BFF suggests that BFF and F2 may have been comparably energetic, at a few $\times$10$^{36}$ erg.
The fact that the bolometric energy estimates generated each of two ways differ from each other by about
a factor of three or so suggests that the energy partition is not constant from flare to flare.
Considering the upper limits to X-ray radiated energy and the \Vfilt-band radiated energy, the ratio
varies from $E_{X}$/$E_{V} \approx$ 28 for BFF, and $\approx$ 20 for F2. 
The BFF event had
significantly more coronal radiation than F2 when considered relative to the optical radiated energy.
The data are too sparse to determine whether there is a relationship between the size of the flare
or other parameters and the energy partition determined. 

\begin{deluxetable}{lll}
\tablewidth{0pt}
\tablenum{4}
\tablecolumns{5}
\tablecaption{Energy Partition in DG~CVn BFF and F2 events\tablenotemark{*}
\label{tbl:epart}
}
\tablehead{ \colhead{Filter/Bandpass} &  \colhead{DG~CVn BFF} &\colhead{DG~CVn F2}  }
\startdata
X-ray (0.3-10 keV) & 4$\times$10$^{35}$ & 9$\times$10$^{35}$ \\
X-ray (0.01-100 keV) & 4.5$-\leq$8$\times$10$^{35}$ & $\leq$10$^{36}$\\
\Vfilt\ & 2.8$\times$10$^{34}$ & 5.2$\times$10$^{34}$ \\
\Rfilt\ & \ldots & 8.5$\times$10$^{34}$ \\
\textit{U}\tablenotemark{1} & \textit{4.7$\times$10$^{34}$} & \textit{8.8$\times$10$^{34}$} \\
\textit{E$_{\rm bol,U}$}\tablenotemark{2} & \textit{4.2$\times$10$^{35}$} & \textit{8.0$\times$10$^{35}$} \\
\textit{E$_{\rm bol,X}$}\tablenotemark{3} & \textit{1.3$\times$10$^{36}$} & \textit{3$\times$10$^{36}$} \\
\enddata
\tablenotetext{*}{Unit for the energies is erg. Italicized numbers are derived; see \S~\ref{sec:epart} for details.}
\tablenotetext{1}{\Ufilt\ band energy derived from \Vfilt\ band energy and E$_{U}$-E$_{V}$ scaling of \citet{hawleypettersen}.}
\tablenotetext{2}{Bolometric radiated energy derived from estimated \Ufilt\ band energy and $E_{U}$-E$_{\rm bol}$
scaling of \citet{ostenwolk2015}.}
\tablenotetext{3}{Bolometric radiated energy derived from 0.3-10 keV X-ray energy and $E_{X}$-E$_{\rm bol}$
scaling of \citet{ostenwolk2015}.}
\end{deluxetable}

\subsubsection{Kinetic Energy\label{sec:ekin}}
\begin{figure}[!h]
\begin{minipage}[c]{0.49\linewidth}
\includegraphics[scale=0.45]{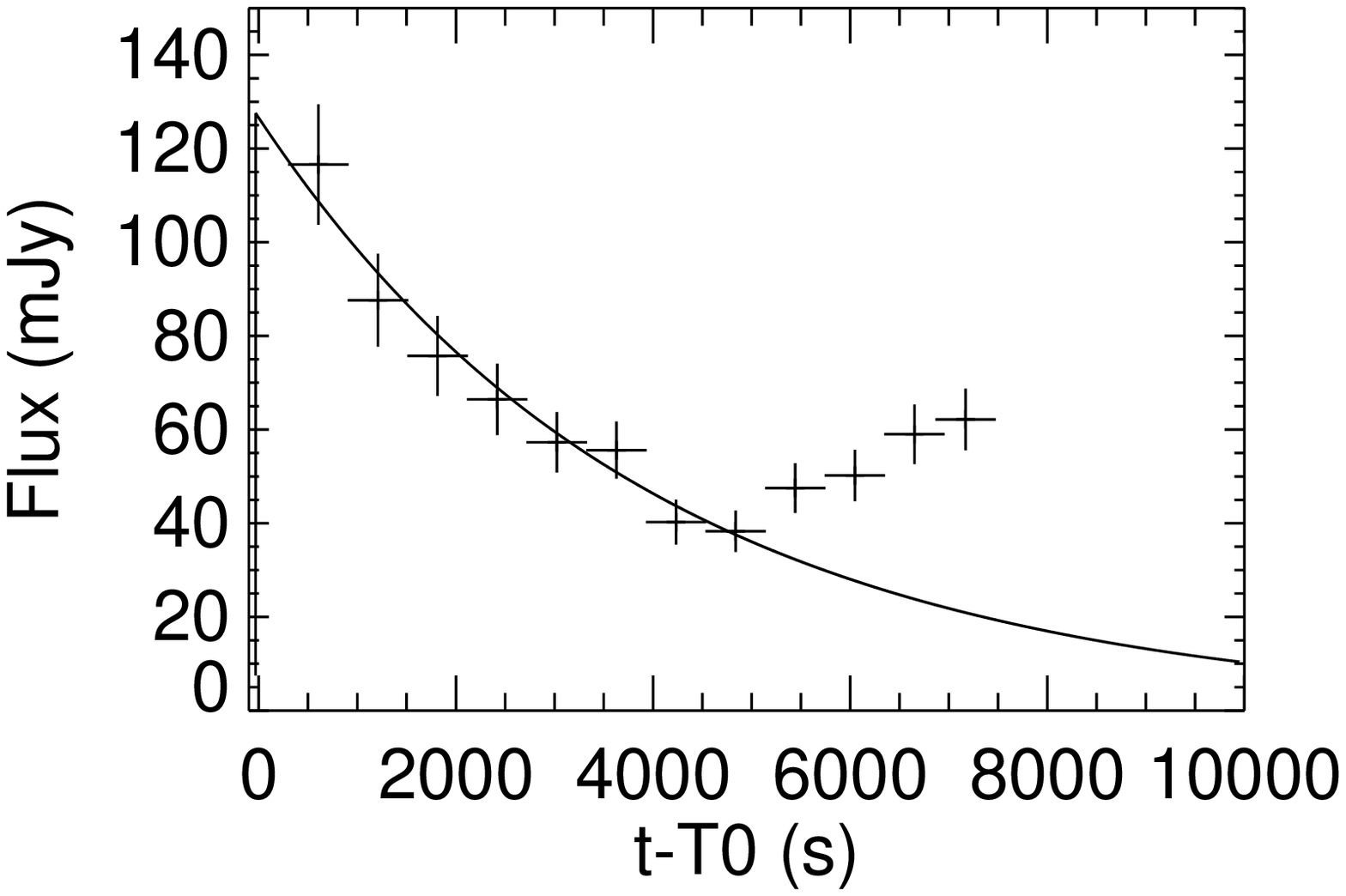}
\end{minipage}
\hfill
\begin{minipage}[c]{0.49\linewidth}
\includegraphics[scale=0.37,angle=90]{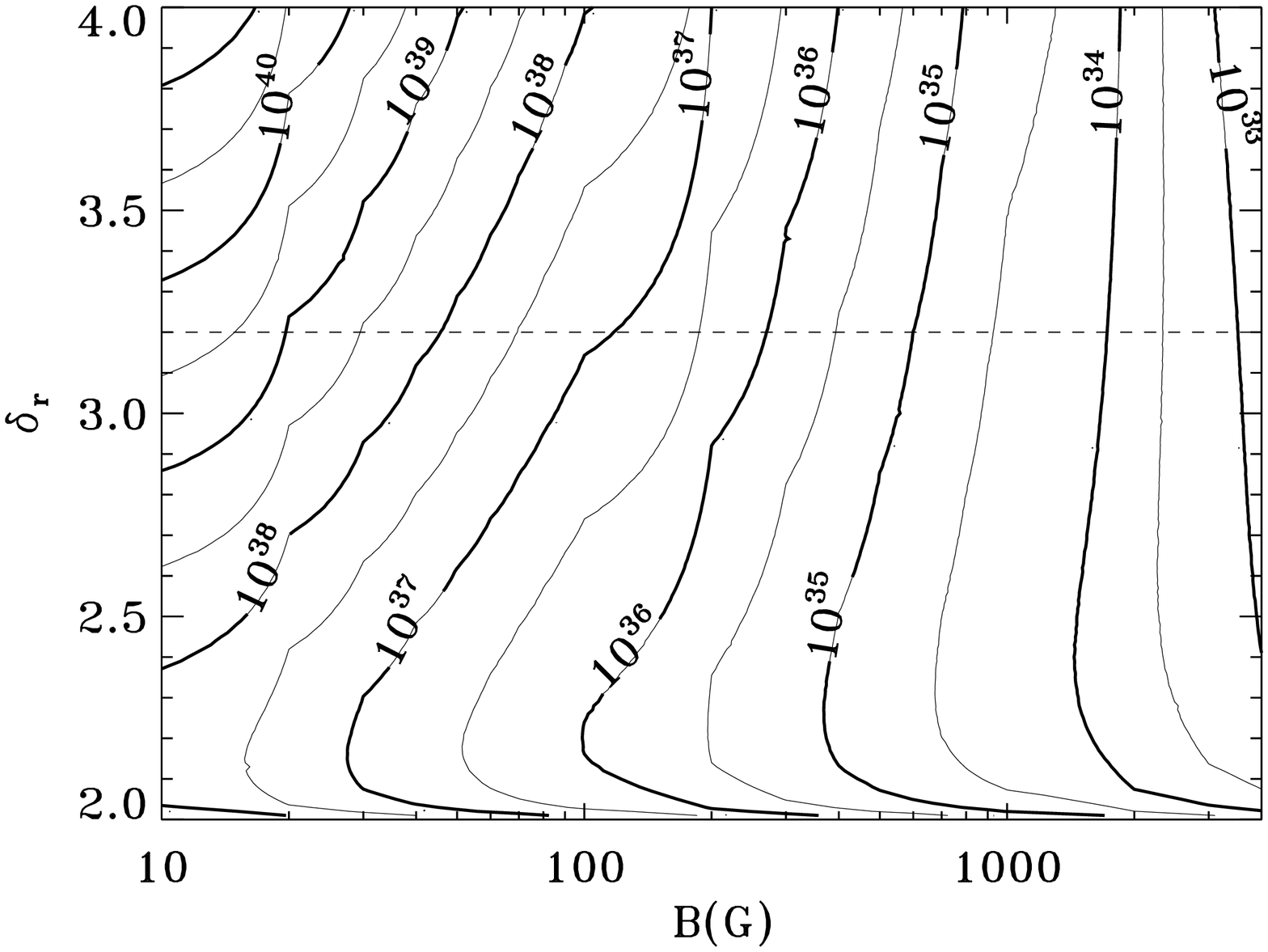}
\end{minipage}
\caption{\textbf{(left)} Radio light curve of BFF, from \citet{fender2015}. Plusses indicate the temporal extent of each bin and the
flux uncertainties. The main event is described as a linear rise and exponential decay, shown in the thick black line.
\textbf{(right)} 
Dependence of kinetic energy in electrons on the index of the electron energy distribution $\delta_{r}$
and the magnetic field strength in the radio-emitting region $B$. Thick lines give values of the kinetic energy
in powers of ten, and thin lines give the value at half-decade intervals.  The dotted line shows the value
of $\delta_{X}$ derived from fitting the X-ray spectra as nonthermal thick-target bremsstrahlung emission,
from \S~\ref{sec:ntfits}.
\label{fig:ami}
}
\end{figure}

\citet{fender2015} presented 15.7 GHz data obtained starting about 6 minutes after the Swift trigger of DG~CVn.
Since the radio flare traces the action of nonthermal particles, these measures constrain
the amount of kinetic energy in the BFF.
We follow the treatment of \citet{smith2005} in estimating the kinetic energy from the radio light curve.
By assuming a spectral energy distribution of the radio emission, and modelling the temporal evolution of
the emission, we can estimate the total radio energy, and hence the kinetic energy. The free
parameters are the magnetic field strength in the radio-emitting source and the distribution of 
the accelerated electrons.
The radio light curve data is taken from \citet{fender2015}, and the portion within $\sim$ 2 hours of the trigger is shown in Figure~\ref{fig:ami};
we use these data to constrain the time profile of the radio flare.
Since the radio data suggest there was a single decline from the peak, we model the time profile of BFF as a
single exponential decay, with a peak at the start time suggested from the start of the \Vfilt\ band burst
reported in \citet{caballero2015}, namely $T0-70$ seconds. 
We assume a fast linear rise to the maximum flux. The decay is fit from the radio light curve,
and is 3980 s; we assume there is a single exponential decay during the decline of the radio flare.
We also assume that the radio spectrum has a spectral shape of the form \\
\begin{eqnarray}
S_{\nu} = A \nu^{\alpha_{1}} \;\;\; \rm for \;\; \nu \le \nu_{\rm pk} \\
S_{\nu} = B \nu^{\alpha_{2}} \;\;\; \rm for \;\; \nu \ge \nu_{\rm pk} 
\end{eqnarray}
where $S_{\nu}$ is the radio flux density, $\nu_{\rm pk}$ is the peak frequency, separating optically thick ($\nu \le \nu_{\rm pk}$) emission with spectral index $\alpha_{1}$ 
from optically thin ($\nu \ge \nu_{\rm pk}$) emission with spectral index $\alpha_{2}$, and A and B are prefactors 
describing the dependence of the radio emission on other parameters.
We assume that the peak frequency
does not change during the decay, although there is evidence from solar and stellar flares
that the peak frequency does change during the impulsive phase \citep{lee2000,osten2005}.
The spectral indices for a homogeneous radio-emitting source, on either side of $\nu_{\rm pk}$ are \\
\begin{eqnarray}
\alpha_{1}& =& 2.5+0.085 \delta_{r} \\
\alpha_{2}&=& 1.22-0.90\delta_{r}
\end{eqnarray}
\citep{dulk1985}, where $\delta_{r}$ is the index of the distribution of nonthermal electrons.
The dependence of the number density of nonthermal electrons with energy and time, $n(E,t)$ is a
separable function and has the form \\
\begin{equation}
n(E,t)=\frac{N(t)(\delta_{r} -1)}{E_{0}} (E/E_{0})^{-\delta_{r}} \;\; 
\end{equation}
where $E$ is the electron energy, $N(t)$ describes the temporal behavior of the number of accelerated particles,
and $E_{0}$ is a cutoff energy, usually taken to be 10 keV \citep{dulk1985}.
Using this formalism, the time evolution seen at $\nu_{AMI}=$15.7 GHz, $F(t)$,  can be applied to all frequencies\\
\begin{eqnarray}
S(\nu,t)& =& F(t) \left(\frac{\nu}{\nu_{AMI}}\right)^{\alpha_{2}} \;\;\; {\rm for \;\; \nu \ge \nu_{\rm pk}}\\
S(\nu,t)& =& \frac{F(t) \nu_{\rm pk}^{(\alpha_{2}-\alpha_{1})}}{\nu_{AMI}^{\alpha_{2}}} \nu^{\alpha_{1}} \;\;\; {\rm for \;\; \nu \le \nu_{\rm pk}}
\end{eqnarray}
where we have assumed $\nu_{pk} < \nu_{AMI}$.
Solar and stellar radio observations show that the peak frequency $\nu_{\rm pk}$ is usually $\sim$10 GHz.
Since DG~CVn is at a known distance, we then convert this to $L_{r}(\nu,t)$ (erg s$^{-1}$ Hz$^{-1}$)
to describe the
temporal and spectral behavior of the flare.

The kinetic energy at a given time is then determined by integrating over the energy dependence from
the lower energy cutoff to infinity \\
\begin{equation}
E_{\rm kin}(t) = N(t)V(t) \frac{\delta_{r}-1}{\delta_{r}-2} E_{0}
\end{equation}
where $V(t)$ is the source volume.  For optically thin emission the flux density can be expressed as \\
\begin{eqnarray}
S(\nu,t)& = &k \nu^{2}/c^{2} \int T_{b}(\nu,t) d\Omega(t) \\
        & = & \frac{\eta (\nu,t)V(t)}{d^{2}} 
\end{eqnarray}
with $k$ Boltzmann's constant, $c$ the speed of light, $T_{b}$ the brightness temperature,
and $d\Omega$ the solid angle subtended by the radio-emitting source. 
The equation can be rewritten using 
$T_{b}= c^{2}/k/\nu^{2} \eta (\nu) L$ for $\tau_{\nu} \ll$1, 
$\eta (\nu)$ is the gyrosynchrotron emissivity, L the length scale of radio-emitting material, and $d$ the stellar distance. 
The radio luminosity can then be expressed as \\
\begin{equation}
L_{r}(\nu,t)= 4 \pi \eta(\nu,t) V(t)
\end{equation}
We use the analytic expressions for the emissivity $\eta$ for X-mode emission from \citet{dulk1985} \\
\begin{equation}
\eta(\nu,t) = 3.3\times10^{-24} 10^{-0.52 \delta_{r}} B N(t) (\sin \theta)^{-0.43+0.65\delta_{r}}\times \left( \frac{\nu}{\nu_{B}} \right)^{1.22-0.9\delta_{r}}
\end{equation}
where $B$ is the magnetic field strength in the radio-emitting source, $\theta$ the angle between the
radio emitting region and the line of sight, $\nu_{B}$ the electron gyrofrequency.
We define $A(\nu)$ to contain the constant and frequency-dependent prefactors, and substitute this into $\eta$
\begin{equation}
\eta(\nu,t) = A(\nu) N(t), \;\;\; \rm{with}
\end{equation}
\begin{equation}
A(\nu) = 3.3\times10^{-24} 10^{-0.52 \delta_{r}} B (\sin \theta)^{-0.43+0.65\delta_{r}}\times \left( \frac{\nu}{\nu_{B}}\right)^{1.22-0.9\delta_{r}}
\end{equation}
where we are assuming that the magnetic field strength in the radio-emitting source does not change appreciably with time.
Then L$_{r}$ can be expressed as \\
\begin{equation}
L_{r}(\nu,t)= 4\pi A(\nu) N(t) V(t) \;\; .
\end{equation}
This can be rearranged so that \\
\begin{equation}
N(t)V(t) = \frac{\int L_{r} (t,\nu) d\nu}{4\pi \int A(\nu) d\nu}
\end{equation}
and \\
\begin{equation}
E_{\rm kin,tot} =  \int N(t) V(t) \frac{\delta_{r}-1}{\delta_{r}-2} E_{0} dt.
\end{equation}
Particles will be depleted and replenished during this time; this can be accounted for in the temporal
variations.
The incompleteness of the observational data, coupled with some of the assumptions made in the analysis, will not render
this a precise estimate of the kinetic energy, but does allow for an order of magnitude estimation, given 
the magnetic field strength in the radio-emitting source and a constraint on the energy distribution of accelerated particles.
Since stellar radio data are usually consistent with a relatively hard radio spectra, we examine
$\delta_{r}$ in the range 2.2$\leq \delta_{r} \leq$ 3.9.
The right panel of Figure~\ref{fig:ami} shows the parametric dependence of the kinetic energy on the unknown values of
magnetic field strength and index of the nonthermal electron distribution.
The implied kinetic energy ranges from very large values, of $\sim$ 10$^{40}$ erg for low values of magnetic field
(tens of Gauss)
and high values of $\delta_{r}$, to 10$^{34}$ erg and less for kiloGauss fields and a range of $\delta_{r}$.
These constraints will be used in Section \ref{sec:bffdisc} to aid in the constraint on the thermal or nonthermal nature of
the X-ray emission in BFF.

\subsection{Coronal Loop Length Determination \label{sec:loop}}
We used the method of \citet{reale1997} to infer a loop length corresponding to the X-ray emission from the
decaying phase of F2.  As described in \citet{osten2010} applied to Swift data, the 
thermodynamic loop decay time can be expressed \citep{serio1991} as \\
\begin{equation}
\tau_{th} = \alpha l/\sqrt{T_{\rm max}}
\end{equation}
where $\alpha$=3.7$\times$10$^{-4}$ cm$^{-1}$ s$^{-1}$K$^{1/2}$, $l$ is the loop half-length in cm,
and $T_{\rm max}$ is the flare maximum temperature (K), \\
\begin{equation}
T_{\rm max} = 0.0261 T_{\rm obs}^{1.244}
\end{equation}
and $T_{\rm obs}$ is the maximum best-fit temperature derived from single temperature fitting
of the data. 
The ratio of the observed exponential light curve decay time $\tau_{LC}$ to the thermodynamic
decay time $\tau_{th}$ can be written as a function which depends on the slope $\zeta$ of the decay
in the log(n$_{e}$-$T_{e}$) plane (or equivalently, $\log(\sqrt{\mathscr{VEM}}-T)$ plane) and other parameters \\
\begin{equation}
\tau_{LC}/\tau_{th} = \frac{c_{a}}{\zeta -\zeta_{a}} + q_{a} = F(\zeta) \;\;\; .
\end{equation}
The parameters fit the functional form above and described in \citet{reale1997};
$q_{a}$, $c_{a}$, and $\zeta_{a}$ are determined for the Swift/XRT instrument
to be $q_{a}$=0.67$\pm$0.33, $c_{a}$=1.81$\pm$0.21, and $\zeta_{a}$=0.1$\pm$0.05 (F. Reale priv. comm.). Combining 
the above expression with the one for the thermodynamic loop decay time, a
relationship between flare maximum temperature, light curve exponential
decay time, and slope in the density-temperature plane can be used to estimate the
flare half-length \\
\begin{equation}
l=\frac{\tau_{LC} \sqrt{T_{\rm max}}}{\alpha F(\zeta)} \;\;,
\label{eqn:loopl}
\end{equation}
valid for $0.4 \leq \zeta\leq 1.9$. 
The errors on plasma parameters quoted in Table~\ref{tbl:f2specfit} are 90\% confidence
intervals, whereas the uncertainties on $q_{a}$, $c_{a}$, and $\zeta_{a}$ are 1$\sigma$ values, so we
recomputed 1$\sigma$ uncertainties for temperature and $\mathscr{VEM}$ to calculate $\zeta$ and coronal loop length
in a consistent fashion.

\begin{figure}[!h]
\includegraphics[scale=0.5,angle=90]{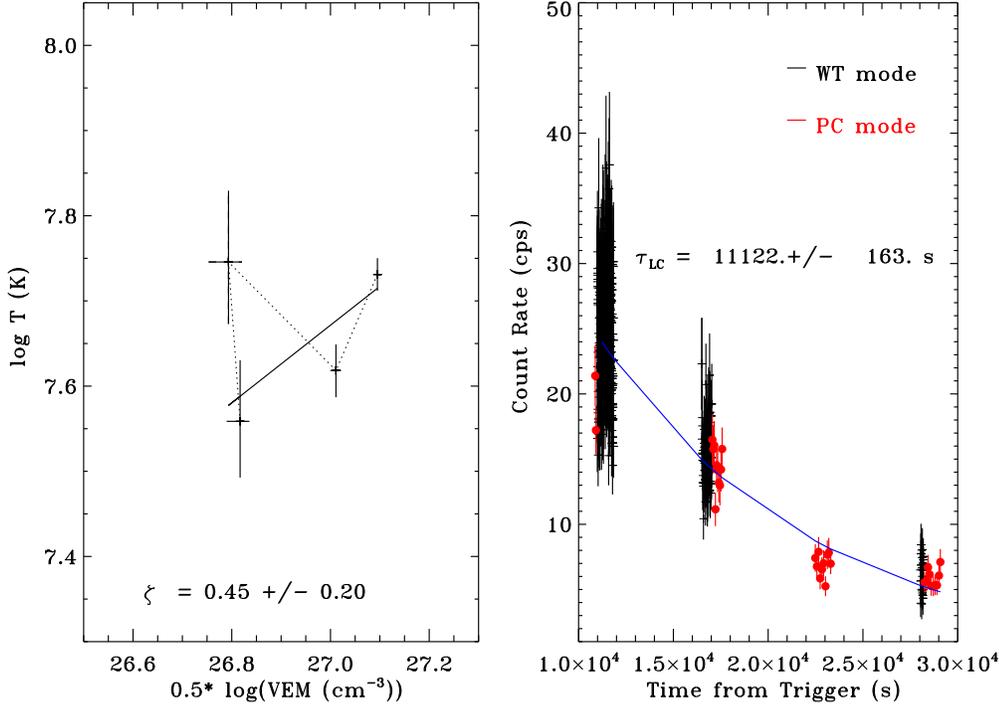}
\caption{
\textbf{(left)} The trend of temperature versus the square root of volume emission measure derived from each 
spectrum, for the F2 event, along with a determination of the slope $\zeta$, given by the solid line.
Dotted line connects data points
in temporal order, starting from upper rightmost point. One $\sigma$ error bars on temperature and $\mathscr{VEM}$ are used in the analysis.
\textbf{(right)} Decay time of F2 from light curve.
\label{fig:zeta}
}
\end{figure}

Figure~\ref{fig:zeta} shows T, $\mathscr{VEM}$ measurements of F2, from which we derive $\zeta=$0.45$\pm$0.20. 
The right panel of Figure~\ref{fig:zeta} shows the XRT light curve during
the decay phase of F2, where an exponential decay time $\tau_{LC}$ is determined to
be 11122$\pm$163 s.
Combining these parameters with the value of T$_{\rm obs}$ from T0+10860:T0+11849 in Table~\ref{tbl:f2specfit}
and using 
Equation~\ref{eqn:loopl}, the loop semi-length $l$ is 5.4$\pm$2.9$\times$10$^{10}$ cm, or
2.0$\pm$1.0 stellar radii. Assuming a circular loop, this suggests a height above the photosphere of $2l/\pi$ or 
1.3$\pm$0.6 R$_{\star}$.

\section{Discussion }
The F2 event seems like a typical example of a superflare: hot plasma is produced and footpoint emission
occurs, displaying the response of the upper stellar atmosphere (the X-ray-emitting corona) as well as the
stellar photosphere (optical photometry) to the deposition of energy from presumably a
magnetic reconnection event. The coronal loop lengths implied by an analysis of the decay phase of F2 are in line with
results from other large stellar flares, which have displayed loop semi-lengths up to 10$^{12}$ cm on young
stars \citep{favata2005,mcclearywolk2011}.  
The interpretation of BFF is more complicated, as the X-ray
spectra show the dominance of either a superhot thermal plasma component, or nonthermal X-ray emission.
Both of these possibilities are extreme. The former would be the hottest spectroscopically confirmed
plasma temperature in a stellar flare, outstripping the T$_{X}\sim$100 MK seen in other superflares
\citep{osten2007,osten2010}. There is only one other claim on nonthermal hard X-ray emission in 
a stellar superflare, that of \citet{osten2007}. Given the close proximity in time of F2 and BFF,
possibly formed in the same active region, we examine the implications of the F2 event
to see what it can tell us about BFF and ultimately, the nature of these extreme stellar superflares.

\subsection{F2 Event \label{sec:f2disc}}
The energetics of F2 in different bands were determined in \S~\ref{sec:Xengy}, \S~\ref{sec:Rengy}, 
and \S~\ref{sec:Vengy},
estimate of the coronal loop length in \S~\ref{sec:loop}, and optical footpoint area in \S~\ref{sec:foot}.
Using these pieces of information, along with other parameters estimated from spectroscopic analysis, we
determine some other parameters for F2 which will be useful in comparison with BFF.
These numbers are tabulated in Table~\ref{tbl:compare}.

For the F2 event, the relative energetics in the \Vfilt\ and \Rfilt\ filters compare favorably with
the Great Flare on the nearby M dwarf AD Leo described by HP91,
even though the overall energetics are about a factor of 20 higher.
In that event, based on their Table 6, 1.8 times more energy was released in the \Rfilt\ band
compared to the \Vfilt\ band. As described in \S~\ref{sec:Rengy} and  \S~\ref{sec:Vengy},
approximately the same energy ratio is observed for F2.
The X-ray energy for F2, estimated in \S~\ref{sec:Xengy}, is an order of magnitude larger still. 
While this is a large number in consideration of the typical radiated flare energies from nearby M dwarfs,
it is not out of line with extremes of activity seen in very young stars. 
\citet{caramazza2007} studied X-ray flares on low mass stars in Orion, and found X-ray flares
in the 0.5-8 keV range with X-ray radiated energies up to 2$\times$10$^{35}$ erg.
\citet{mcclearywolk2011} studied high-contrast flares in young stars, and found X-ray flares up to
about 10$^{37}$ erg.


\subsubsection{Determination of Other Parameters for F2}

The X-ray decay analysis in \S~\ref{sec:loop} gives a loop semi-length of $l=2.0\pm1.0$ R$_{\star}$.
Based on the discussion in \S~\ref{sec:foot},  we use constraints on the flare footpoint
area from \Vfilt\ and \Rfilt\ band photometry for F2 obtained assuming a color temperature
T$_{\rm BB}$ of 6000 K, or X$_{\rm F2,V}$(T$_{\rm BB}$=6000 K)=0.14.
The total flaring area is 
$A_{fl}=X_{fl} \pi R_{\star}^{2}$=2$\pi$R$_{\rm foot}^{2}$.
These two numbers constrain the value $\alpha=R_{\rm foot}/(2l)$, assuming a single columnar loop with semi-length $l$ 
as derived above and two footpoints contributing to the total optical area; 
the flare area becomes \\
\begin{equation}
A_{fl}=X_{\rm F2,V} \pi R_{\star}^{2} = 2 \pi R_{\rm foot}^{2},
\end{equation}
with R$_{\rm foot}$ the radius of one footpoint,
and the value $\alpha=R_{\rm foot}/(2l)$ can then be determined
to be 0.07 using the vlaue of $X_{\rm F2,V}$ evaluated for T$_{BB}$=6000 K.
This is independent of the number of loops, as long as the loop length $l$ doesn't change when
generalized to N flaring loops. 
We can couple this with a simple picture of the emitting region as a loop (or N flaring loops), where
the volume emission measure can be expressed as \\
\begin{equation}
\mathscr{VEM} = n_{e}^{2} \pi R_{\rm foot}^{2} 2l \;\;.
\end{equation}
This can be rearranged to give \\
\begin{equation}
\label{eqn:den}
n_{e} = \left[  \frac{\mathscr{VEM} \alpha}{\pi R_{\star}^{3}} \left(\frac{X_{F2,V}}{2}    \right)^{-3/2}   \right]^{1/2}
\end{equation}
and with the $\mathscr{VEM}$ constrained from the spectrum at peak of F2, and $\alpha$ from the combination of 
the white-light event and coronal loop length, the coronal electron density is constrained to be 3$\times$10$^{11}$
cm$^{-3}$.
The quantity $\alpha$ in Equation~\ref{eqn:den} is calculated for $N=1$.
Note that this approach can be applied to $N$ flaring loops, and the electron density is unchanged.
With $n_{e}$ and $T_{X}$, the strength of the magnetic field required to confine the flaring coronal plasma is then \\
\begin{equation}
\label{eqn:bconf}
B_{\rm conf} = \sqrt{8\pi n_{e}kT_{X}}
\end{equation}
with $k$ Boltzmann's constant; evaluating this, we derive $B_{\rm conf}$ of $\sim$230 G for F2.

The \Vfilt-band measurements at the peak of F2 as well as X-ray measurements constrain the flux ratio,
useful for a comparison of the relative brightening of photospheric and coronal emissions, respectively.
At T0+9709$\pm$63 s, the Johnson \Vfilt\ magnitude is 9.83.  This gives a flux density of 4.3$\times$10$^{-13}$
erg cm$^{-2}$ s$^{-1}$ $\AA^{-1}$, or integrated flux over the \Vfilt\ filter bandpass of 3.6$\times$10$^{-10}$
erg cm$^{-2}$ s$^{-1}$, using a FWHM of 836 \AA. From Table~\ref{tbl:f2specfit} the 0.3-10 keV flux from 
the nearest interval,
T0+10860:T0+11849 s, is 6.59$\times$10$^{-10}$ erg cm$^{-2}$ s$^{-1}$.
We use the flux estimated in the 0.01-100 keV energy range, 8$\times$10$^{-10}$ erg cm$^{-2}$ s$^{-1}$,
for a flux ratio $f_{X}/f_{V}$
of 2.2.

\begin{figure}
\includegraphics[scale=0.8]{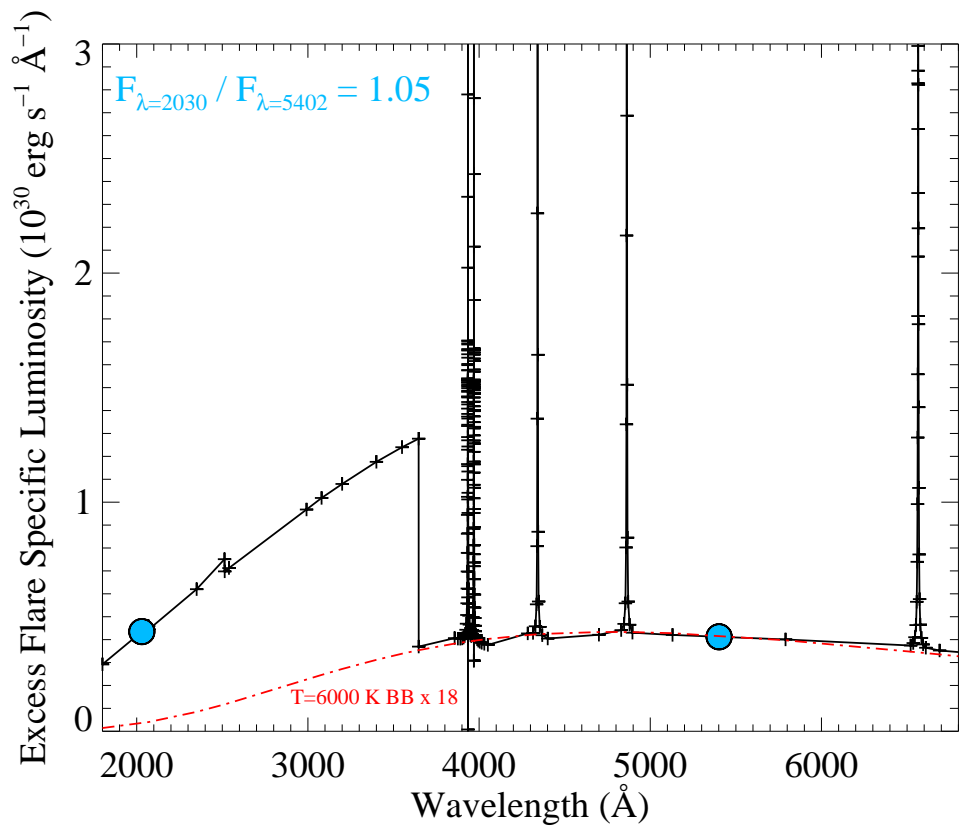}
\caption{Excess flare specific luminosity calculated using the RADYN code as described in \citet{kowalski2015};
black plusses and solid lines are the RHD calculations. Red dash-dot line is the blackbody spectrum at T$_{BB}$=6000K.
Blue circles are the midpoints of the \uvwfilt\ and \Vfilt\ filters, respectively, and are used to estimate a flux ratio
to be compared against that observed in F2.
\label{fig:app}}
\end{figure}

{\bf Is X$\sim$0.14 from a 6000 K blackbody reasonable?  A preliminary multithread modeling approach to F2 uses the 
F13 beam heated atmosphere from K13, which were calculated with the RADYN code
\citep{radynref}.  If we assume that the emission during F2 can be modeled by a superposition of 
impulsively heated loops (new kernels using the ``average burst'' spectrum 
\citep[Table 1 of ][; $F_{\rm kernel}$ here]{kowalski2015})
and decay phase emission from previously heated loops (F13 gradual decay spectrum at $t=4$s in Table~1 of K15; $F_{\rm decay}$) with area coverage 25x the kernel emission, then we obtain a broad 
band spectrum that is generally consistent with the  coarse Swift UV and optical colors (\uvwfilt/\Vfilt$\sim$1 compared
to the observations $\sim$0.8-0.9).  We can estimate an areal 
coverage using an actual RHD spectrum
\begin{equation}
F_{\rm flare} = 25 * X_{\rm kernel} * F_{\rm decay} + X_{\rm kernel} * F_{\rm kernel}
\end{equation}
where $F_{\rm kernel}$ is the surface flux of the F13 model averaged over its evolution;
$F_{\rm decay}$ is the surface flux of the F13 during the gradual phase.

For the peak of F2, $X_{\rm kernel}$ = 0.008 and 25 * $X_{\rm kernel}$ = 0.2  which is similar to 0.14 for T=6000 K.
For this modeling, X(T=10,000K) = 0.023 is justified for a nonthermal interpretation and X(T=6000 K) = 0.14 is 
justified for 
a thermal (decaying loop) interpretation.  In \S 4.1.1, the best area to use is likely the T=6000 K area,
although the RHD model (at $t=4$s) that is used to represent the decay emission from previously
heated loops is far shorter than the decay times obtained from the X-ray light curves in Section 3.
In Figure~\ref{fig:app} we show the flare specific luminosity (L$_{\rm flare}$=$\pi R_{\rm star}^{2}$F$_{\rm flare}$)
for the multithread model from RADYN.
}

\citet{aschwanden2008} following on earlier work by \citet{shibatayokoyama1999} found a striking similarity between
power-law dependences of flare peak volume emission measure against temperature for a sample of solar and stellar 
flares. The index of the scaling was the same for solar and stellar flares, 
even while the stellar flare temperatures were generally hotter.
The trend, for stars, was $\mathscr{VEM}_{p} = 10^{50.8} \left( \frac{T_{p}}{10 MK} \right) ^{4.5\pm0.4} \;\; cm^{-3}$
with T$_{p}$ being the flare peak temperature and $\mathscr{VEM}_{p}$ the peak volume emission measure of the flare.
Using the peak X-ray temperature of F2 from spectroscopic fitting in \S~\ref{sec:f2}, namely 48 MK,
the emission measure expected from this scaling relation is 7$\times$10$^{53}$ cm$^{-3}$, 
so only a factor of 2.3 below the observed
peak emission measure of 1.64$\times$10$^{54}$ cm$^{-3}$.

\subsection{Interpretation of BFF \label{sec:bffdisc}}

Analysis of the BFF event shows that from approximately T0-30 until T0+328 seconds,
the X-ray spectra reveal evidence of either a high temperature plasma, larger than any seen in previous large stellar
flares, or a nonthermal thick-target bremsstrahlung emission. The comparison is especially
good in the $\sim$200 s where both XRT and BAT spectra are available: 
the XRT+BAT spectrum of BFF is remarkably featureless. 
Both models are essentially indistinguishable based on the  statistics of the model fits to the data, and here we turn to 
supporting data to aid in our interpretation.
We use the supporting data available for BFF, namely the \Vfilt\ band measurements which constrain the
flare footpoint area, and the decay of a radio flare starting about 7 minutes after T0
\citep[reported in ][]{fender2015}, from which we have estimated
$E_{\rm kin}$ in \S~\ref{sec:ekin}. 

X-ray and optical measurements of BFF early after the trigger enable a determination of the ratio
of coronal to photospheric emissions.  
The \Vfilt\ peak magnitude from \citet{caballero2015} is $\sim$7, giving a flux of about 5.83$\times$10$^{-12}$
erg cm$^{-2}$ s$^{-1}$ \AA$^{-1}$.
Then the estimated \Vfilt\ filter flux at this time
is 4.9$\times$10$^{-9}$ erg cm$^{-2}$ s$^{-1}$.
From Table~\ref{tbl:specfit} the 0.3-10 keV flux
for BFF at a time near the \Vfilt\ measurement is 4.5$\times$10$^{-9}$ erg cm$^{-2}$ s$^{-1}$
from T0-30:T0+72, extrapolated from the best-fit model in the 14-100 keV energy range.  
Because of the large fraction of X-ray flux emitted in the 14-100 keV range, we use the total (0.01-100 keV)
energy range estimated to be 9.3$\times$10$^{-9}$ erg cm$^{-2}$ s$^{-1}$.
This leads to a flux ratio $f_{X}/f_{V}$ of 1.9, 
similar to that obtained near the peak of F2.

A lower limit for the estimated 0.3-10 keV X-ray radiated energy of BFF assuming an exponential decay is about 4$\times$10$^{35}$ erg, 
less than half the radiated energy from F2. The amount of energy radiated in the \Vfilt\ filter bandpass
is about an order of magnitude less than this.  
\S~\ref{sec:epart} discusses the energy partition within BFF and F2; both appear to be X-ray luminous
compared to expectations from scaling of optical flare energy to bolometric radiated energy
from \citet{ostenwolk2015}.
The integrated \Vfilt\ band energy from the EV~Lac superflare described
in \citet{osten2010} could not be calculated due to insufficient data, but we note that
the large enhancement flare on the very low mass star described in \citet{stelzer2006} had nearly equal amounts
of radiated energy in the X-ray and \Vfilt\ filter bandpasses (but overall lower integrated
values, at $\sim$3$\times$10$^{32}$ erg).



\subsubsection{A Nonthermal Interpretation for BFF}
Table~\ref{tbl:specfit} lists the best-fit parameters for the trigger spectrum and $\approx$ 200 seconds where both
XRT and BAT spectra were obtained, using a nonthermal thick-target bremsstrahlung model for
the spectral fitting.  The free parameters in the spectral fitting
are the index $\delta_{X}$ of the accelerated electrons and the total power in the electron beam. 

We have one constraint on the kinetic energy of the accelerated particles
involved in the event by multiplying the power from each spectral segment
by the integration time for that segment. From the parameters in Table~\ref{tbl:specfit}, this is 
a staggering 5$\times$10$^{39}$ erg, and is a lower limit, since there is no X-ray data for a large portion of the
event, and the total power depends on the value of the low energy cutoff, as described in \S 3.1.1.
Estimation of the kinetic energy also proceeded from the radio light curve in Section~\ref{sec:ekin}; we
use the value of $\delta_{X}$ from the X-ray spectral fits to determine plausible values of kinetic energy.
Solar flare data shows that the nonthermal electrons producing nonthermal hard X-ray emission tend to be
less energetic than those producing the radio gyrosynchrotron emission, but we assume $\delta_{X}=\delta_{r}$
for simplicity.
With that substitution, the kinetic energy then becomes a function of the magnetic field strength in the radio-emitting
source, according to Figure~\ref{fig:ami}. 
Matching the lower limit on kinetic energy implied by a nonthermal interpretation of the X-ray spectrum
with the estimated kinetic energy inferred from analysis of the radio flare
requires a very low magnetic field strength, of order 20 G or less in the radio-emitting source.

 The peak power of the electron beam, derived from spectral fitting, is 3$\times$10$^{37}$ erg s$^{-1}$.
The \Vfilt\ band measurement at the peak of BFF gives a constraint on the footpoints of the flaring loop to be
X$_{\rm BFF,V}=$0.375, or an area of 9$\times$10$^{20}$ cm$^{2}$. This implies a beam flux of 10$^{16}$ erg cm$^{-2}$ s$^{-1}$, which
 is about four orders of magnitude larger than the largest beam fluxes investigated for solar flares
\citep[$>$5$\times$10$^{12}$ erg s$^{-1}$ cm$^{-2}$][]{krucker2011}.  
For an F16 beam the drift speed of the return current would have to be the speed of light and still the beam
would not be neutralized, therefore resulting in strong magnetic fields.
Such a large beam flux seems physically 
implausible, as it would require treatment of a return current and violation of fundamental assumptions, and
we do not consider it to be a viable interpretation.
Additionally, \citet{caballero2015} argue that the time delay between the optical \Vfilt\ band burst 
and the Swift trigger is evidence of the Neupert effect, which would be difficult to envisage if
the X-ray emission was entirely nonthermal.

\subsubsection{A thermal interpretation for BFF}
The standard flare decay analyses for the X-ray emission done for F2 will not work for BFF, 
because the temperature is not changing appreciably over the 200 second timescales over which we have X-ray data.
Our insight into BFF is guided by analysis of the F2 event, for which
we see $T(t)$, $\mathscr{VEM}(t)$, and from which we can infer coronal loop length, and a ratio of the
radius of the loop (from white-light footpoints) to the coronal loop length. 
For the BFF event, \Vfilt\ band data give us the flare footpoint area, and by assuming that the same value of $\alpha$
applies to BFF as well as F2, we can estimate the 
loop length for BFF. For $\alpha=0.07$, and the value of $X_{BFF,V}=0.375$ for a black-body
temperature of 10$^{4}$K, we infer a coronal loop semi-length of 
3.2 R$_{\star}$ or maximum height of 2.0 R$_{\star}$.
Tying these parameters together with the peak $\mathscr{VEM}$ for the thermal model from Table~\ref{tbl:specfit}
using equations ~\ref{eqn:den} and ~\ref{eqn:bconf},
we derive $n_{e}$ of 3$\times$10$^{11}$ cm$^{-3}$, and a 
$B_{\rm conf}$ of 580 G.
These numbers are similar to what we derive for F2, and given that the energy partition between
X-rays and \Vfilt\ band appears to be similar, the likely case is a thermal plasma.

Table~\ref{tbl:compare} compares key parameters of the two flares considered here on DG~CVn, as well as the superflare
on the nearby EV~Lac described in \citet{osten2010}. The energy comparison is restricted to the energy bands with the
most temporal coverage: in EV~Lac and BFF there was significant HXR emission in the initial stages of the flare.
Note that the estimation of $f_{V}$ for EV~Lac, proceeding using the same  steps as described above for F2 event
of DG CVn, yields an integrated \Vfilt\ filter flux of 6.15$\times$10$^{-9}$ erg cm$^{-2}$ s$^{-1}$, for a value of
f$_{X}/f_{V}$ of 4.0. The peak X-ray luminosity relative to bolometric luminosity 
was calculated over the expanded energy range of 0.01-100 keV
to account for the majority of the radiative losses of the hot coronal plasma; the value for EV~Lac was taken 
using parameters in the first line of Table~2 in \citet{osten2010} and calculated on this larger energy range.
Using the scaling relationship between flare temperature and emission measure established
for solar flares and a sample of stellar flares,
for the DG~CVn flare BFF
we would expect a $\mathscr{VEM}$ nearly 2400 times larger than observed,  and for the EV~Lac peak flare 
a factor of 88 larger than observed (see Table~\ref{tbl:compare}). 
This may be seen as problematic for the thermal interpretation, however there have been previous
suggestions of a departure from this behavior at the highest stellar flare temperatures previously
observed.
\cite{getman2008} suggested that superhot flares may turn over
in this relationship, based on 
inferring flare temperatures using a median energy analysis of flares from
young stars, and suggested that even at temperatures in excess of 100 MK the $\mathscr{VEM}$ should be between 10$^{54}$
and 10$^{55}$ cm$^{-3}$.  

\begin{figure}
\includegraphics[scale=0.8]{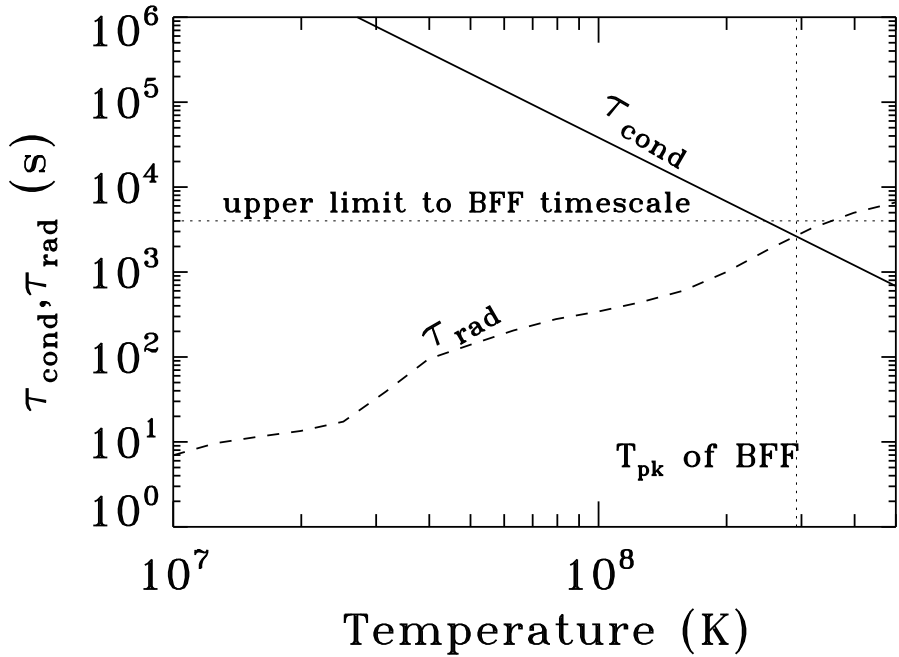}
\caption{Dependence of conductive and radiative loss times versus temperature for the BFF event, using electron
densities and length scales calculated in analysis. The location of the peak temperature of BFF derived from
X-ray spectral fitting, as well as the upper limit to the timescale of the event, are also shown, and are consistent
with the timescale and temperature where the two loss mechanisms are approximately equal.
\label{fig:loss}}
\end{figure}

The high temperatures suggest that the plasma will lose its energy by conductive losses on a relatively
fast timescale.  
The timescale on which the plasma will lose energy by radiative losses \\
\begin{equation}
\tau_{\rm rad} = \frac{3k_{B}T_{e}}{n_{e} \psi(T_{e})}
\end{equation}
depends on the electron density $n_{e}$, electron temperature T$_{e}$, Boltzmann's constant $k_{B}$, and
the radiative loss function $\psi(T_{e})$.  The radiative losses for a collisionally ionized
plasma are evaluated by summing the contributions from
line and continuum radiation at each temperature tabulated in the Astrophysical Plasma Emission Database \citep{apecref},
similar to what was done in \citet{osten2006}.
The timescale on which the plasma will lose energy by conductive losses is \\
\begin{equation}
\tau_{\rm cond} = \frac{3 n_{e} k_{B} l^{2}}{\kappa T_{e}^{5/2}}
\end{equation}
where $l$ is the length scale and $\kappa$ is the Spitzer conductivity coefficient \citep[=8.8 $\times$ 10$^{-7}$ ergs cm$^{-1}$ 
s$^{-1}$ K$^{-7/2}$][]{spitzer1962}.
Using values for electron density and length scale derived from analyses above, we determine the dependence of the two timescales
on electron temperature and compare with the duration and peak temperature of BFF. Figure~\ref{fig:loss}
displays the results.  It is curious that the location at which the two timescales are approximately equal
is close to the peak temperature of BFF, and the value of the timescales are similar to the upper limit
given to the event duration of BFF from the sparse data.

Given the extreme parameters of BFF, we can also examine the ratio of the relaxation time of the plasma to the conductive cooling time.
In order for such a hot temperature plasma to be
observed, the relaxation time should not be larger than the timescale on which the plasma lose energy by 
conduction.  The ratio of
the thermal relaxation time of the plasma to the timescale for conductive cooling is\\
\begin{equation}
\label{eqn:relax}
\tau_{\rm relax}/\tau_{\rm cond} = 2 \frac{T_{8}^{4}}{n_{11}^{2}L_{9}^{2}}
\end{equation}
\citep[see discussion in ][]{benzbook}, where $T_{8}$ is the temperature in units of 10$^{8}$ K,
$n_{11}$ is the electron density in units of 10$^{11}$ cm$^{-3}$, and $L_{9}$ is the
loop length in units of 10$^{9}$ cm.
Evaluating Equation~\ref{eqn:relax} for the values appropriate for this flare, we find $\tau_{\rm relax}/\tau_{\rm cond}$
to be 2$\times$10$^{-4}$.
This demonstrates that the plasma does have time to relax to the observed thermal temperature.

The total kinetic energy in BFF cannot be constrained independently for a thermal interpretation of BFF; the
results from \S3.3.4 show it to depend on $\delta$ and B.
However, an estimate for a lower limit to the total radiated X-ray energy for BFF is 4$\times$10$^{35}$ erg.
Studies of the global energetics of large solar flares \citep{emslie2012} show that the total radiation
from soft X-ray emitting plasma is comparable to or slightly smaller than the energy in flare electrons
accelerated to energies greater than 20 keV.  If we assume that the energy partition is similar for BFF, then 
a rough equipartition between the radiated X-ray energy and the energy in accelerated electrons
would suggest, via Figure~\ref{fig:ami},  magnetic field strengths in the radio-emitting plasma of 
several hundred Gauss to about 1 kiloGauss. This is consistent with the field strengths derived 
above independently from equipartition between the gas pressure and magnetic pressure.

If we take the thermal interpretation as the more physically plausible explanation, given the 
constraints from the multi-wavelength observations, then we can still ask the question of what
signature of nonthermal electrons might be expected to appear in the hard X-ray spectral range.
Nonthermal particles propagate in a collisionless plasma.  A lower limit to the particle energy
required to cross a propagation path with length $L$ across a density $n_{e}$ is given by 
setting the propagation time of accelerated electrons equal to the collisional deflection time
\citep{aschwanden2002}\\
\begin{equation}
E \geq 20 \sqrt{L_{9} n_{11} \left( \frac{0.7}{\cos \alpha} \right)} \;\;\; keV
\end{equation}
where $L_{9}$ is the loop length in units of 10$^{9}$ cm, $n_{11}$ is the electron density in
units of 10$^{11}$ cm$^{-3}$, and $\alpha$ is the pitch angle. For the parameters
in Table~\ref{tbl:compare} the minimum energy is 580 keV; this analysis suggests that 
the accelerated particles filling the entire flare loop
would produce nonthermal hard X-ray emission at energies above this 
to be  potentially observable.

\begin{deluxetable}{llll}
\tablewidth{0pt}
\tablenum{5}
\tablecolumns{4}
\tablecaption{Comparison of DG~CVn BFF and F2 events with EV~Lac superflare \label{tbl:compare}
}
\tablehead{ \colhead{} &  \colhead{DG~CVn BFF} &\colhead{DG~CVn F2} & \colhead{EV~Lac\tablenotemark{*}} }
\startdata
stellar spectral type & M4V & M4V & M3V \\
 dist. (pc) & 18 & 18 & 5 \\
stellar age (MY) & 30 & 30 & 30$-$300 \\
Peak Temperature T$_{X}$\tablenotemark{1} (10$^{6}$K) &290  &54 &  139\\
Peak $\mathscr{VEM}$\tablenotemark{2} (10$^{54}$ cm$^{-3}$) &9 &1.55 & 6.3 \\
$\mathscr{VEM}_{\rm expected}$\tablenotemark{3} (10$^{54}$ cm$^{-3}$) &2400 &1.2 &88 \\
L$_{X,peak,0.01-100 keV}$/L$_{\rm bol}$ &4.8 &0.4 & 3.5\\
Integrated Energy (0.3-10 keV) (10$^{34}$ erg) & 40 &90 & 5.8 \\
Integrated Energy (\Vfilt\ band)  (10$^{34}$ erg)   &2.8   &5.2 &$\ldots$ \\
Footpoint Fractional Area X$_{\rm fl}(10^{4}K)$ &0.375$^{V}$  &0.023$^{V}$ &$>$0.03$^{v}$ \\
$f_{X}/f_{V}$ & 1.9 & 2.2 & 4.0 \\
Duration$_{X}$ (hr) & 2.2 & 6.4 & 1.7 \\
FWHM$_{V}$ (s) & 20 & 3600 & \ldots \\
Loop Semi-length (R$_{\star}$) & 3.2  &2.0$\pm$1.0 &0.37$\pm$0.07 \\
n$_{e}$ (10$^{11}$ cm$^{-3}$) &3 &3&30 \\
B$_{\rm conf}$ (G) &580 &230 &1100 \\
\enddata
\tablenotetext{*}{Data for EV~Lac taken from Osten et al. (2010)}
\tablenotetext{1}{Peak Temp. from BAT+XRT} 
\tablenotetext{2}{Peak $\mathscr{VEM}$ obtained at different time from peak T}
\tablenotetext{3}{$\mathscr{VEM}$ expected using $T-\mathscr{VEM}$ scaling of \citet{aschwanden2008}}
\tablenotetext{v}{: area taken from Swift \vfilt\ band; $^{V}$: area taken from \Vfilt\ band}
\end{deluxetable}

\subsection{Implications}
The two large flares studied on DG~CVn in this
paper are both an order of magnitude larger than the individual flares on nearby M dwarfs previously studied in
detail, and they also eclipse the radiated energies of the largest flares seen on much younger stars 
not amenable to detailed study.  It is remarkable that in one dataset we have possibly the top two
most energetic X-ray flares from a low mass star that have been detected to date.
The large stellar flares on M dwarfs previously studied in detail have tended to be much lower in energy
and amplitude. This is understandable as the frequency of occurrence of large flares declines with both
increasing energy and peak luminosity.  
\citet{caramazza2007} reported on X-ray flares
occurring on $\sim$1 MY old solar-mass and low-mass stars, with the low-mass stars having flares with radiated
energies in the 0.3-8 keV bandpass of up to $\sim$2$\times$10$^{35}$ erg. Due to the larger distances of the
low mass flaring stars in their sample (at the distance of the Orion Nebula Cluster, or $\approx$ 450 pc),
detailed study of the X-ray flares was not possible, and there was no accompanying multi-wavelength
information. 

We do not have solid constraints on the frequency of events of this large size. 
\citet{tsuboi2014} reported approximately 4 stellar flares from active M dwarfs
with energies in the range 10$^{35}$-10$^{36}$ which are not upper limits, from four
years' worth of monitoring of the hard X-ray sky with the MAXI/GSC instrument, which
suggests an upper bound occurrence rate of roughly one flare per year per star.  
If we use flare frequency distributions for an active M dwarf and extrapolate to these energies, then we can get a 
lower bound on occurrence rate. 
\citet{lme1976} calculated flare frequencies for a sample of nearby active M dwarfs using
integrated \Ufilt-band energies.  From Table~\ref{tbl:epart}, we have estimated the \Ufilt-band energies
for BFF and F2, respectively.  Using the three single flare stars in \citet{lme1976} which had the
largest flare energies in that paper, namely YZ~CMi, EQ~Peg, and EV~Lac, we calculate the
expected occurrence rate for flares exceeding 4.7$\times$10$^{34}$ erg (E$_{U}$ for BFF) to be 
once every (68, 388, 69) days for (YZ~CMi, EQ~Peg, and EV~Lac), respectively. 
If we consider the two events combined to be a single large eruptive event, then the occurrence rate of such
energetic events, which together total E$_{U}$=1.3$\times$10$^{35}$ erg, is once every (141, 1080, 140)
days for the flare frequency distribution parameters from YZ~CMi, EQ~Peg, and EV~Lac. These estimates
vary by a factor of 10 from each other, and reflect not only the uncertainty in flare to flare differences
in flare frequency distributions, but also uncertainty in the behavior of the flare frequency distribution itself; 
namely whether the occurrence rate of flares at such high energies continues to follow a power-law distribution.
These estimates are also about a factor of 3 from the estimate using
\citet{tsuboi2014} indicating general agreement at about that level. 

These events are also far larger than the event studied in detail to determine the likely astrobiological
effects of stellar flares on close-in terrestrial exoplanets, and have coverage in both the UV/optical and 
X-ray bandpasses. 
Given the estimated occurrence rate, they will be an important contributor in shaping
the radiation and particle environment around an M dwarf in which extrasolar planets will be forming and existing.
\citet{segura2010} utilized 
UV-optical observations of the Great Flare on AD Leo, which had an integrated energy in the 1200-8000 \AA\
range of $\sim$10$^{34}$ erg. They used scalings between UV radiated energy and X-ray energy, and a further
scaling of X-ray flux to proton flux, to model the response of a terrestrial atmosphere to the impingement of
the UV flare photons only or the flare photons along with the MeV energy protons.  
We note that the comparison of the \Rfilt\ band radiated energy of F2 on DG~CVn to the Great Flare on AD~Leo
in our Figure~\ref{fig:Renergy} reveals F2 to be a  factor of $\approx$ 20 larger. 
The energetic protons which were modelled removed the ozone layer, on timescales of about 2 years, and the recovery time
of the planetary atmosphere was a few decades. These effects coupled with the esimated occurrence rate of
even larger events suggests that there might be a permanent erosion of the ozone layer.  
One critical open question in this area is whether the scalings observed in solar eruptive events between 
photons and particles holds during stellar events.
A more recent paper on the effects of stellar flares on exoplanetary atmospheres \citep{venot2016}
determined that planets around very active stars would likely never achieve a steady state due to
the frequent photon bombardment of the exoplanetary atmosphere from stellar flares. 

\citet{smith2004} considered the transport of ionizing radiation in terrestrial exoplanet atmospheres,
and found that while a thick atmosphere can protect the planetary surface from incident
X-rays and $\gamma$-rays, up to 4\% of the incident ionizing radiation received at the surface
in the 2000-3200 \AA\ wavelength range comes from atmospheric transmission and reprocessing of 
the high energy radiation. As the two flares reported here have peak X-ray luminosities more than several hundred
times larger than the quiescent values, any irradiation of a planetary atmosphere would increase
temporarily by the same large factor. 
The effect of such large variations in the stellar ionizing flux may be significant,
especially early in the life of the star and planetary system.
The previous studies examining the impact of stellar flares on exoplanet atmospheres,
like \citet{segura2010} and \citet{venot2016}, 
used only UV observations, ignoring the potential increase in UV 
emission incident on the planet due to high energy photons in the manner described in \citet{smith2004}.
Hence they are underestimates at best of the impact of these flares on exoplanet atmospheric chemistry.

The flaring star which produced these immense energetic releases is not solitary, which
raises the question of whether the companion could have any impact on the existence of these
superflares. 
The DG~CVn system is a binary,
with the two stars separated by $\sim$4 au. At their young age the components are not yet tidally synchronized,
precluding an origin in increased magnetic activity from tidal interaction akin to that seen in RS~CVn or BY~Dra systems.
Recent studies of M dwarf-white dwarf binary systems suggest that even non-interacting close binaries ($<10$ au separation)
may have a higher flare rate
than single stars \citep{morgan2016}, although \citet{stern1995} found no correlation between X-ray luminosity 
and orbital period for spectroscopic binaries in the Hyades with periods greater than 10 days.
Whether this can be related to dynamical interaction within a circumbinary disk compared to the evolution of a circumstellar disk
is speculative. However, since single M dwarf flare stars are also capable of producing stellar superflares with 
roughly similar characteristics (Table~\ref{tbl:compare})
this suggests that the binarity is not a strong
factor.

\section{Conclusions}
We presented a detailed study of two of the most energetic flare events seen on a young low mass star.
In addition to measurements made for each flare event, we used the properties of the second flare F2
to infer some of the properties for BFF.
The results confirmed the basic flare scenario for hyperactive stars as for solar flares, and revealed
evidence of departures of trends between the temperatures and emission measures of the highest
temperature stellar flares compared with lower temperature solar flares.
The object, DG~CVn, has been relatively uncharacterized for its flaring and extreme magnetic activity
and we hope that this report will spur additional studies. 
Based on the flare properties described in this paper, 
we expect the existence of
very strong magnetic fields
in the photosphere.  Starspot modelling should confirm the nature of starspot sizes 
implied by the flare footpoint modelling. Uncertainties in the rotation period and $v\sin i$
mentioned in the introduction are likely the result of the previously unrecognized binary nature
of the system.

While X-ray flares from stars are commonly known, observations with Swift have revealed that
stellar flares can be bright enough to trigger the BAT with their intense hard X-ray ($>$15 keV)
emission. These events reveal the nature of magnetic reconnection processes occurring in a
regime vastly different from the Sun, yet exhibiting continuity with solar events. 
Supporting
data from both space- and ground-based observatories enable more constraints on the 
extremes of energetics and plasma parameters. 
In contrast with the claim of nonthermal emission from the superflare on II~Peg
reported by \cite{osten2007}, for the DG~CVn event the possibility of a nonthermal interpretation
is confronted with constraints on kinetic energy and photospheric flare area provided
by radio and optical observations, respectively.
Since the nonthermal interpretation is disfavored in the DG~CVn flares because of constraints from the radio
and optical data, the II~Peg nonthermal interpretation is in doubt.

The extreme nature of the flare temperature of BFF, coupled with results from other extreme flares,
suggest that the scaling between solar and stellar flare temperatures and emission measures
exhibits a flattening at high temperatures.  The opportunity these flares 
present to confirm this flattening by using spectroscopically derived temperatures is important
and may reveal departures from canonical solar flare behavior.

Planets around M dwarfs will likely experience millions of these kinds of superflares during their infancy.
This pair of well-studied flares on M dwarfs should be used to provide updated constraints
on the impact of flare radiation on close-in terrestrial exoplanets.
This confirms the conclusion reached for EV Lac that the ``habitable zone" $\sim$ 0.1 AU from a young M dwarf star is 
likely 
inimicable to life: the flare peak luminosity in the GOES (1.5 - 8 keV) band would be equivalent to an X60,000,000 
flare. If the energetic proton fluxes and coronal mass ejection energies scale with the radiated flare energy, the 
impact upon the atmosphere and magnetosphere of any hypothetical terrestrial  planet would be catastrophic.

\acknowledgements
This work made use of data supplied by the UK Swift Science Data Centre at the
University of Leicester.
This publication makes use of data products from the Wide-field Infrared Survey Explorer, which is a joint project of the University of California, Los Angeles, and the Jet Propulsion Laboratory/California Institute of Technology, funded by the National Aeronautics and Space Administration. 
SRO also acknowledges the support of the Spanish Ministry, Project Number AYA2012-39727-C03-01.
This publication makes use of data products from the Wide-field Infrared Survey Explorer, which is a joint project of the University of California, Los Angeles, and the Jet Propulsion Laboratory/California Institute of Technology, funded by the National Aeronautics and Space Administration. 
We acknowledge the support from the Swift project (N. Gehrels) and Swift schedulers at Penn State, which
enabled the acquisition of this wonderful data set.
RAO and AK acknowledge fruitful discussions at ISSI in Bern with the Energy Transformation in Solar and Stellar Flares team
during the preparation of this manuscript.


\begin{thebibliography}{}
\expandafter\ifx\csname natexlab\endcsname\relax\def\natexlab#1{#1}\fi

\bibitem[{{Aschwanden}(2002)}]{aschwanden2002}
{Aschwanden}, M.~J. 2002, \ssr, 101, 1

\bibitem[{{Aschwanden} {et~al.}(2008){Aschwanden}, {Stern}, \&
  {G{\"u}del}}]{aschwanden2008}
{Aschwanden}, M.~J., {Stern}, R.~A., \& {G{\"u}del}, M. 2008, \apj, 672, 659

\bibitem[{{Barthelmy} {et~al.}(2005){Barthelmy}, {Barbier}, {Cummings},
  {Fenimore}, {Gehrels}, {Hullinger}, {Krimm}, {Markwardt}, {Palmer},
  {Parsons}, {Sato}, {Suzuki}, {Takahashi}, {Tashiro}, \& {Tueller}}]{batref}
{Barthelmy}, S.~D., {Barbier}, L.~M., {Cummings}, J.~R., {et~al.} 2005, Space
  Science Reviews, 120, 143

\bibitem[{{Beers} {et~al.}(1994){Beers}, {Bestman}, \& {Wilhelm}}]{beers1994}
{Beers}, T.~C., {Bestman}, W., \& {Wilhelm}, R. 1994, \aj, 108, 268

\bibitem[{{Benz}(2002)}]{benzbook}
{Benz}, A. 2002, {Plasma Astrophysics, second edition} (Plasma
  Astrophysics.~Kinetic Processes in Solar and Stellar Coronae, second
  edition.~By A.~Benz, Institute of Astronomy, ETH Z{\"u}rich, Switzerland.~
  Astrophysics and Space Science Library, Vol.~279, Kluwer Academic Publishers,
  Dordrecht, 2002.)

\bibitem[{{Benz} \& {G{\"u}del}(2010)}]{benzgudel2010}
{Benz}, A.~O., \& {G{\"u}del}, M. 2010, \araa, 48, 241

\bibitem[{{Beuzit} {et~al.}(2004){Beuzit}, {S{\'e}gransan}, {Forveille},
  {Udry}, {Delfosse}, {Mayor}, {Perrier}, {Hainaut}, {Roddier}, {Roddier}, \&
  {Mart{\'{\i}}n}}]{beuzit2004}
{Beuzit}, J.-L., {S{\'e}gransan}, D., {Forveille}, T., {et~al.} 2004, \aap,
  425, 997

\bibitem[{{Breeveld} {et~al.}(2011){Breeveld}, {Landsman}, {Holland}, {Roming},
  {Kuin}, \& {Page}}]{breeveld2011}
{Breeveld}, A.~A., {Landsman}, W., {Holland}, S.~T., {et~al.} 2011, in American
  Institute of Physics Conference Series, Vol. 1358, American Institute of
  Physics Conference Series, ed. J.~E. {McEnery}, J.~L. {Racusin}, \&
  N.~{Gehrels}, 373--376

\bibitem[{{Brickhouse} {et~al.}(2010){Brickhouse}, {Cranmer}, {Dupree}, {Luna},
  \& {Wolk}}]{brickhouse2010}
{Brickhouse}, N.~S., {Cranmer}, S.~R., {Dupree}, A.~K., {Luna}, G.~J.~M., \&
  {Wolk}, S. 2010, \apj, 710, 1835

\bibitem[{{Burrows} {et~al.}(2005){Burrows}, {Hill}, {Nousek}, {Kennea},
  {Wells}, {Osborne}, {Abbey}, {Beardmore}, {Mukerjee}, {Short}, {Chincarini},
  {Campana}, {Citterio}, {Moretti}, {Pagani}, {Tagliaferri}, {Giommi},
  {Capalbi}, {Tamburelli}, {Angelini}, {Cusumano}, {Br{\"a}uninger}, {Burkert},
  \& {Hartner}}]{xrtref}
{Burrows}, D.~N., {Hill}, J.~E., {Nousek}, J.~A., {et~al.} 2005, Space Science
  Reviews, 120, 165

\bibitem[{{Caballero-Garc{\'{\i}}a} {et~al.}(2015){Caballero-Garc{\'{\i}}a},
  {{\v S}imon}, {Jel{\'{\i}}nek}, {Castro-Tirado}, {Cwiek}, {Claret}, {Opiela},
  {{\.Z}arnecki}, {Gorosabel}, {Oates}, {Cunniffe}, {Jeong}, {Hudec},
  {Sokolov}, {Makarov}, {Tello}, {Lara-Gil}, {Kub{\'a}nek}, {Guziy}, {Bai},
  {Fan}, {Wang}, \& {Park}}]{caballero2015}
{Caballero-Garc{\'{\i}}a}, M.~D., {{\v S}imon}, V., {Jel{\'{\i}}nek}, M.,
  {et~al.} 2015, \mnras, 452, 4195

\bibitem[{{Caramazza} {et~al.}(2007){Caramazza}, {Flaccomio}, {Micela},
  {Reale}, {Wolk}, \& {Feigelson}}]{caramazza2007}
{Caramazza}, M., {Flaccomio}, E., {Micela}, G., {et~al.} 2007, \aap, 471, 645

\bibitem[{{Carlsson} \& {Stein}(1997)}]{radynref}
{Carlsson}, M., \& {Stein}, R.~F. 1997, \apj, 481, 500

\bibitem[{{Chrastina} \& {Hroch}(2008)}]{munipack}
{Chrastina}, M., \& {Hroch}, F. 2008, Open European Journal on Variable Stars,
  95, 21

\bibitem[{{Cutri} \& {et al.}(2013)}]{allwise}
{Cutri}, R.~M., \& {et al.} 2013, VizieR Online Data Catalog, 2328

\bibitem[{{Demory} {et~al.}(2009){Demory}, {S{\'e}gransan}, {Forveille},
  {Queloz}, {Beuzit}, {Delfosse}, {di Folco}, {Kervella}, {Le Bouquin},
  {Perrier}, {Benisty}, {Duvert}, {Hofmann}, {Lopez}, \& {Petrov}}]{demory2009}
{Demory}, B.-O., {S{\'e}gransan}, D., {Forveille}, T., {et~al.} 2009, \aap,
  505, 205

\bibitem[{{Dennis} \& {Schwartz}(1989)}]{dennis1989}
{Dennis}, B.~R., \& {Schwartz}, R.~A. 1989, \solphys, 121, 75

\bibitem[{{Drake} {et~al.}(2014){Drake}, {Osten}, {Page}, {Kennea}, {Oates},
  {Krimm}, \& {Gehrels}}]{dgcvnATel}
{Drake}, S., {Osten}, R., {Page}, K.~L., {et~al.} 2014, The Astronomer's
  Telegram, 6121, 1

\bibitem[{{Dulk}(1985)}]{dulk1985}
{Dulk}, G.~A. 1985, \araa, 23, 169

\bibitem[{{Emslie} {et~al.}(2012){Emslie}, {Dennis}, {Shih}, {Chamberlin},
  {Mewaldt}, {Moore}, {Share}, {Vourlidas}, \& {Welsch}}]{emslie2012}
{Emslie}, A.~G., {Dennis}, B.~R., {Shih}, A.~Y., {et~al.} 2012, \apj, 759, 71

\bibitem[{{Evans} {et~al.}(2007){Evans}, {Beardmore}, {Page}, {Tyler},
  {Osborne}, {Goad}, {O'Brien}, {Vetere}, {Racusin}, {Morris}, {Burrows},
  {Capalbi}, {Perri}, {Gehrels}, \& {Romano}}]{evans2007}
{Evans}, P.~A., {Beardmore}, A.~P., {Page}, K.~L., {et~al.} 2007, \aap, 469,
  379

\bibitem[{{Evans} {et~al.}(2009){Evans}, {Beardmore}, {Page}, {Osborne},
  {O'Brien}, {Willingale}, {Starling}, {Burrows}, {Godet}, {Vetere}, {Racusin},
  {Goad}, {Wiersema}, {Angelini}, {Capalbi}, {Chincarini}, {Gehrels}, {Kennea},
  {Margutti}, {Morris}, {Mountford}, {Pagani}, {Perri}, {Romano}, \&
  {Tanvir}}]{evans2009}
---. 2009, \mnras, 397, 1177

\bibitem[{{Favata} {et~al.}(2005){Favata}, {Flaccomio}, {Reale}, {Micela},
  {Sciortino}, {Shang}, {Stassun}, \& {Feigelson}}]{favata2005}
{Favata}, F., {Flaccomio}, E., {Reale}, F., {et~al.} 2005, \apjs, 160, 469

\bibitem[{{Fender} {et~al.}(2015){Fender}, {Anderson}, {Osten}, {Staley},
  {Rumsey}, {Grainge}, \& {Saunders}}]{fender2015}
{Fender}, R.~P., {Anderson}, G.~E., {Osten}, R., {et~al.} 2015, \mnras, 446,
  L66

\bibitem[{{Gershberg}(1972)}]{Gershberg1972}
{Gershberg}, R.~E. 1972, \apss, 19, 75

\bibitem[{{Getman} {et~al.}(2008){Getman}, {Feigelson}, {Broos}, {Micela}, \&
  {Garmire}}]{getman2008}
{Getman}, K.~V., {Feigelson}, E.~D., {Broos}, P.~S., {Micela}, G., \&
  {Garmire}, G.~P. 2008, \apj, 688, 418

\bibitem[{{Gizis} {et~al.}(2002){Gizis}, {Reid}, \& {Hawley}}]{gizis2002}
{Gizis}, J.~E., {Reid}, I.~N., \& {Hawley}, S.~L. 2002, \aj, 123, 3356

\bibitem[{{Hawley} \& {Fisher}(1992)}]{Hawley1992}
{Hawley}, S.~L., \& {Fisher}, G.~H. 1992, \apjs, 78, 565

\bibitem[{{Hawley} \& {Pettersen}(1991)}]{hawleypettersen}
{Hawley}, S.~L., \& {Pettersen}, B.~R. 1991, \apj, 378, 725

\bibitem[{{Hawley} {et~al.}(2003){Hawley}, {Allred}, {Johns-Krull}, {Fisher},
  {Abbett}, {Alekseev}, {Avgoloupis}, {Deustua}, {Gunn}, {Seiradakis}, {Sirk},
  \& {Valenti}}]{Hawley2003}
{Hawley}, S.~L., {Allred}, J.~C., {Johns-Krull}, C.~M., {et~al.} 2003, \apj,
  597, 535

\bibitem[{{Helfand} {et~al.}(1999){Helfand}, {Schnee}, {Becker}, {White}, \&
  {McMahon}}]{helfand1999}
{Helfand}, D.~J., {Schnee}, S., {Becker}, R.~H., {White}, R.~L., \& {McMahon},
  R.~G. 1999, \aj, 117, 1568

\bibitem[{{H{\"u}nsch} {et~al.}(1999){H{\"u}nsch}, {Schmitt}, {Sterzik}, \&
  {Voges}}]{hunsch1999}
{H{\"u}nsch}, M., {Schmitt}, J.~H.~M.~M., {Sterzik}, M.~F., \& {Voges}, W.
  1999, \aaps, 135, 319

\bibitem[{{Kontar} {et~al.}(2008){Kontar}, {Dickson}, \& {Ka{\v
  s}parov{\'a}}}]{kontar2008}
{Kontar}, E.~P., {Dickson}, E., \& {Ka{\v s}parov{\'a}}, J. 2008, \solphys,
  252, 139

\bibitem[{{Kowalski} {et~al.}(2015){Kowalski}, {Hawley}, {Carlsson}, {Allred},
  {Uitenbroek}, {Osten}, \& {Holman}}]{kowalski2015}
{Kowalski}, A.~F., {Hawley}, S.~L., {Carlsson}, M., {et~al.} 2015, \solphys,
  290, 3487

\bibitem[{{Kowalski} {et~al.}(2010){Kowalski}, {Hawley}, {Holtzman},
  {Wisniewski}, \& {Hilton}}]{Kowalski2010}
{Kowalski}, A.~F., {Hawley}, S.~L., {Holtzman}, J.~A., {Wisniewski}, J.~P., \&
  {Hilton}, E.~J. 2010, \apjl, 714, L98

\bibitem[{{Kowalski} {et~al.}(2012){Kowalski}, {Hawley}, {Holtzman},
  {Wisniewski}, \& {Hilton}}]{Kowalski2012}
---. 2012, \solphys, 277, 21

\bibitem[{{Kowalski} {et~al.}(2013){Kowalski}, {Hawley}, {Wisniewski}, {Osten},
  {Hilton}, {Holtzman}, {Schmidt}, \& {Davenport}}]{Kowalski2013}
{Kowalski}, A.~F., {Hawley}, S.~L., {Wisniewski}, J.~P., {et~al.} 2013, \apjs,
  207, 15

\bibitem[{{Krucker} {et~al.}(2011){Krucker}, {Hudson}, {Jeffrey}, {Battaglia},
  {Kontar}, {Benz}, {Csillaghy}, \& {Lin}}]{krucker2011}
{Krucker}, S., {Hudson}, H.~S., {Jeffrey}, N.~L.~S., {et~al.} 2011, \apj, 739,
  96

\bibitem[{{Kuerster} \& {Schmitt}(1996)}]{cftucflare}
{Kuerster}, M., \& {Schmitt}, J.~H.~M.~M. 1996, \aap, 311, 211

\bibitem[{{Lacy} {et~al.}(1976){Lacy}, {Moffett}, \& {Evans}}]{lme1976}
{Lacy}, C.~H., {Moffett}, T.~J., \& {Evans}, D.~S. 1976, \apjs, 30, 85

\bibitem[{{Lee} \& {Gary}(2000)}]{lee2000}
{Lee}, J., \& {Gary}, D.~E. 2000, \apj, 543, 457

\bibitem[{{Lin}(2011)}]{lin2011}
{Lin}, R.~P. 2011, \ssr, 159, 421

\bibitem[{{Malamut} {et~al.}(2014){Malamut}, {Redfield}, {Linsky}, {Wood}, \&
  {Ayres}}]{malamut2014}
{Malamut}, C., {Redfield}, S., {Linsky}, J.~L., {Wood}, B.~E., \& {Ayres},
  T.~R. 2014, \apj, 787, 75

\bibitem[{{Mann} {et~al.}(2015){Mann}, {Feiden}, {Gaidos}, {Boyajian}, \& {von
  Braun}}]{mann2015}
{Mann}, A.~W., {Feiden}, G.~A., {Gaidos}, E., {Boyajian}, T., \& {von Braun},
  K. 2015, \apj, 804, 64

\bibitem[{{McCleary} \& {Wolk}(2011)}]{mcclearywolk2011}
{McCleary}, J.~E., \& {Wolk}, S.~J. 2011, \aj, 141, 201

\bibitem[{{Meibom} {et~al.}(2007){Meibom}, {Mathieu}, \&
  {Stassun}}]{meibom2007}
{Meibom}, S., {Mathieu}, R.~D., \& {Stassun}, K.~G. 2007, \apjl, 665, L155

\bibitem[{{Mohanty} \& {Basri}(2003)}]{mohantybasri2003}
{Mohanty}, S., \& {Basri}, G. 2003, \apj, 583, 451

\bibitem[{{Morgan} {et~al.}(2016){Morgan}, {West}, \& {Becker}}]{morgan2016}
{Morgan}, D.~P., {West}, A.~A., \& {Becker}, A.~C. 2016, \aj, 151, 114

\bibitem[{{Newton} {et~al.}(2015){Newton}, {Charbonneau}, {Irwin}, \&
  {Mann}}]{newton2015}
{Newton}, E.~R., {Charbonneau}, D., {Irwin}, J., \& {Mann}, A.~W. 2015, \apj,
  800, 85

\bibitem[{{Osten} {et~al.}(2007){Osten}, {Drake}, {Tueller}, {Cummings},
  {Perri}, {Moretti}, \& {Covino}}]{osten2007}
{Osten}, R.~A., {Drake}, S., {Tueller}, J., {et~al.} 2007, \apj, 654, 1052

\bibitem[{{Osten} {et~al.}(2006){Osten}, {Hawley}, {Allred}, {Johns-Krull},
  {Brown}, \& {Harper}}]{osten2006}
{Osten}, R.~A., {Hawley}, S.~L., {Allred}, J., {et~al.} 2006, \apj, 647, 1349

\bibitem[{{Osten} {et~al.}(2005){Osten}, {Hawley}, {Allred}, {Johns-Krull}, \&
  {Roark}}]{osten2005}
{Osten}, R.~A., {Hawley}, S.~L., {Allred}, J.~C., {Johns-Krull}, C.~M., \&
  {Roark}, C. 2005, \apj, 621, 398

\bibitem[{{Osten} \& {Wolk}(2015)}]{ostenwolk2015}
{Osten}, R.~A., \& {Wolk}, S.~J. 2015, \apj, 809, 79

\bibitem[{{Osten} {et~al.}(2010){Osten}, {Godet}, {Drake}, {Tueller},
  {Cummings}, {Krimm}, {Pye}, {Pal'shin}, {Golenetskii}, {Reale}, {Oates},
  {Page}, \& {Melandri}}]{osten2010}
{Osten}, R.~A., {Godet}, O., {Drake}, S., {et~al.} 2010, \apj, 721, 785

\bibitem[{{Pagani} {et~al.}(2011){Pagani}, {Beardmore}, {Abbey}, {Mountford},
  {Osborne}, {Capalbi}, {Perri}, {Angelini}, {Burrows}, {Campana}, {Cusumano},
  {Evans}, {Kennea}, {Moretti}, {Page}, \& {Starling}}]{pagani2011}
{Pagani}, C., {Beardmore}, A.~P., {Abbey}, A.~F., {et~al.} 2011, \aap, 534, A20

\bibitem[{{Page} {et~al.}(2013){Page}, {Kuin}, {Breeveld}, {Hancock},
  {Holland}, {Marshall}, {Oates}, {Roming}, {Siegel}, {Smith}, {Carter}, {De
  Pasquale}, {Symeonidis}, {Yershov}, \& {Beardmore}}]{page2013}
{Page}, M.~J., {Kuin}, N.~P.~M., {Breeveld}, A.~A., {et~al.} 2013, \mnras, 436,
  1684

\bibitem[{{Parker}(1988)}]{parker1988}
{Parker}, E.~N. 1988, \apj, 330, 474

\bibitem[{{Reale} {et~al.}(1997){Reale}, {Micela}, {Peres}, {Betta}, \&
  {Serio}}]{reale1997}
{Reale}, F., {Micela}, G., {Peres}, G., {Betta}, R., \& {Serio}, S. 1997,
  \memsai, 68, 1103

\bibitem[{{Riedel} {et~al.}(2014){Riedel}, {Finch}, {Henry}, {Subasavage},
  {Jao}, {Malo}, {Rodriguez}, {White}, {Gies}, {Dieterich}, {Winters},
  {Davison}, {Nelan}, {Blunt}, {Cruz}, {Rice}, \& {Ianna}}]{riedel2014}
{Riedel}, A.~R., {Finch}, C.~T., {Henry}, T.~J., {et~al.} 2014, \aj, 147, 85

\bibitem[{{Robb}(1994)}]{robb1994}
{Robb}, R. 1994, in Astronomical Society of the Pacific Conference Series,
  Vol.~55, Optical Astronomy from the Earth and Moon, ed. D.~M. {Pyper} \&
  R.~J. {Angione}, 246

\bibitem[{{Roming} {et~al.}(2005){Roming}, {Kennedy}, {Mason}, {Nousek}, {Ahr},
  {Bingham}, {Broos}, {Carter}, {Hancock}, {Huckle}, {Hunsberger}, {Kawakami},
  {Killough}, {Koch}, {McLelland}, {Smith}, {Smith}, {Soto}, {Boyd},
  {Breeveld}, {Holland}, {Ivanushkina}, {Pryzby}, {Still}, \& {Stock}}]{roming}
{Roming}, P.~W.~A., {Kennedy}, T.~E., {Mason}, K.~O., {et~al.} 2005, Space
  Science Reviews, 120, 95

\bibitem[{{Schmidt} {et~al.}(2014){Schmidt}, {Prieto}, {Stanek}, {Shappee},
  {Morrell}, {Bardalez Gagliuffi}, {Kochanek}, {Jencson}, {Holoien}, {Basu},
  {Beacom}, {Szczygie{\l}}, {Pojmanski}, {Brimacombe}, {Dubberley}, {Elphick},
  {Foale}, {Hawkins}, {Mullins}, {Rosing}, {Ross}, \& {Walker}}]{Schmidt2014}
{Schmidt}, S.~J., {Prieto}, J.~L., {Stanek}, K.~Z., {et~al.} 2014, \apjl, 781,
  L24

\bibitem[{{Segura} {et~al.}(2010){Segura}, {Walkowicz}, {Meadows}, {Kasting},
  \& {Hawley}}]{segura2010}
{Segura}, A., {Walkowicz}, L.~M., {Meadows}, V., {Kasting}, J., \& {Hawley}, S.
  2010, Astrobiology, 10, 751

\bibitem[{{Serio} {et~al.}(1991){Serio}, {Reale}, {Jakimiec}, {Sylwester}, \&
  {Sylwester}}]{serio1991}
{Serio}, S., {Reale}, F., {Jakimiec}, J., {Sylwester}, B., \& {Sylwester}, J.
  1991, \aap, 241, 197

\bibitem[{{Sharykin} {et~al.}(2015){Sharykin}, {Struminskii}, \&
  {Zimovets}}]{hotsolarflares}
{Sharykin}, I.~N., {Struminskii}, A.~B., \& {Zimovets}, I.~V. 2015, Astronomy
  Letters, 41, 53

\bibitem[{{Shibata} \& {Yokoyama}(1999)}]{shibatayokoyama1999}
{Shibata}, K., \& {Yokoyama}, T. 1999, \apjl, 526, L49

\bibitem[{{Smith} {et~al.}(2004){Smith}, {Scalo}, \& {Wheeler}}]{smith2004}
{Smith}, D.~S., {Scalo}, J., \& {Wheeler}, J.~C. 2004, \icarus, 171, 229

\bibitem[{{Smith} {et~al.}(2005){Smith}, {G{\"u}del}, \& {Audard}}]{smith2005}
{Smith}, K., {G{\"u}del}, M., \& {Audard}, M. 2005, \aap, 436, 241

\bibitem[{{Smith} {et~al.}(2001){Smith}, {Brickhouse}, {Liedahl}, \&
  {Raymond}}]{apecref}
{Smith}, R.~K., {Brickhouse}, N.~S., {Liedahl}, D.~A., \& {Raymond}, J.~C.
  2001, \apjl, 556, L91

\bibitem[{{Spitzer}(1962)}]{spitzer1962}
{Spitzer}, L. 1962, {Physics of Fully Ionized Gases}

\bibitem[{{Stassun} {et~al.}(2006){Stassun}, {van den Berg}, {Feigelson}, \&
  {Flaccomio}}]{stassun2006}
{Stassun}, K.~G., {van den Berg}, M., {Feigelson}, E., \& {Flaccomio}, E. 2006,
  \apj, 649, 914

\bibitem[{{Stelzer} {et~al.}(2006){Stelzer}, {Schmitt}, {Micela}, \&
  {Liefke}}]{stelzer2006}
{Stelzer}, B., {Schmitt}, J.~H.~M.~M., {Micela}, G., \& {Liefke}, C. 2006,
  \aap, 460, L35

\bibitem[{{Stern} {et~al.}(1995){Stern}, {Schmitt}, \& {Kahabka}}]{stern1995}
{Stern}, R.~A., {Schmitt}, J.~H.~M.~M., \& {Kahabka}, P.~T. 1995, \apj, 448,
  683

\bibitem[{{Tsuboi} {et~al.}(2014){Tsuboi}, {Higa}, {Yamazaki}, \& {MAXI
  Team}}]{tsuboi2014}
{Tsuboi}, Y., {Higa}, M., {Yamazaki}, K., \& {MAXI Team}. 2014, in Suzaku-MAXI
  2014: Expanding the Frontiers of the X-ray Universe, ed. M.~{Ishida},
  R.~{Petre}, \& K.~{Mitsuda}, 138

\bibitem[{{Venot} {et~al.}(2016){Venot}, {Rocchetto}, {Carl}, {Hashim}, \&
  {Decin}}]{venot2016}
{Venot}, O., {Rocchetto}, M., {Carl}, S., {Hashim}, A., \& {Decin}, L. 2016,
  ArXiv e-prints, arXiv:1607.08147

\end{thebibliography}

\end{document}